\documentclass[preprint]{aastex}
\usepackage{psfig}
\usepackage{natbib}

\newcommand{\nn}{\nonumber \\}
\newcommand{\ls}{\mathrel{\raise1.16pt\hbox{$<$}\kern-7.0pt 
\lower3.06pt\hbox{{$\scriptstyle \sim$}}}}         
\newcommand{\gs}{\mathrel{\raise1.16pt\hbox{$>$}\kern-7.0pt 
\lower3.06pt\hbox{{$\scriptstyle \sim$}}}}         

\long\def\comment#1{}

\def\fun#1#2{\lower3.6pt\vbox{\baselineskip0pt\lineskip.9pt
  \ialign{$\mathsurround=0pt#1\hfil##\hfil$\crcr#2\crcr\sim\crcr}}}
\def\lap{\mathrel{\mathpalette\fun <}}
\def\gap{\mathrel{\mathpalette\fun >}}
\newcommand{\ba}{\begin{eqnarray}}
\newcommand{\ea}{\end{eqnarray}}
\newcommand{\be}{\begin{equation}}
\newcommand{\ee}{\end{equation}}

\begin{document}
\title{The abundance of high-redshift objects as a probe of 
non--Gaussian initial conditions}
\author{Sabino Matarrese\altaffilmark{1,2}, Licia Verde\altaffilmark{3}
and Raul Jimenez\altaffilmark{3}}
\altaffiltext{1}{Dipartimento di Fisica "Galileo
Galilei", via Marzolo 8, I-35131  Padova,
Italy. (matarrese@pd.infn.it)}
\altaffiltext{2}{Max-Planck-Institut f\"ur Astrophysik, 
Karl-Schwarzschild-Strasse 1, D-85748 Garching, Germany. 
(sabino@mpa-garching.mpg.de)}
\altaffiltext{3}{Institute for Astronomy, University of Edinburgh, 
Blackford Hill, Edinburgh EH9 3HJ, UK. (lv@roe.ac.uk, raul@roe.ac.uk)}

\begin{abstract}
The observed abundance of high-redshift galaxies and clusters contains
precious information about the properties of the initial perturbations.  We
present a method to compute analytically the number density of objects as a
function of mass and redshift for a range of physically motivated non-Gaussian
models. In these models the non-Gaussianity can be dialed from zero and is
assumed to be small.  We compute the probability density function for the {\it
smoothed} dark matter density field and we extend the Press and Schechter
approach to {\it mildly} non-Gaussian density fields.  The abundance of
high-redshift objects can be directly related to the non-Gaussianity parameter
and thus to the physical processes that generated deviations from the Gaussian
behaviour. Even a skewness parameter of order $0.1$ implies a dramatic change
in the predicted abundance of $z\gap 1$ objects.  Observations from NGST and
X-ray satellites (XMM) can be used to accurately measure the amount of non-Gaussianity in the primordial density field.
\end{abstract}


\section{Introduction} 
In the standard inflationary model for triggering structure formation in the
Universe, there are precise predictions for the properties of the initial
fluctuations: they are adiabatic and follow a nearly-Gaussian distribution,
with deviations from Gaussianity which are calculable, small and generally
dependent on the specific inflationary model
\citep{FRS93,GLMM94,Gangui94,Wang99,Ganguimartin99}. Because of the smallness
of such deviations from the Gaussian behaviour, in most theoretical
predictions one simply assumes that the primordial density field has {\it
exactly} random phases.  Consequently, intrinsic temperature fluctuations in
the Cosmic Microwave Background (CMB) are commonly treated as being Gaussian
and the same assumption is made in most analyses of Large-Scale Structure
(LSS) of the Universe.  Besides this `standard' nearly-Gaussian model for
generating cosmological structures, based on the amplification of quantum
fluctuations of the same scalar field which drives the inflationary dynamics,
there exist alternative models for the origin of fluctuations which predict
stronger deviations from the random-phase paradigm. Still within the context
of inflation multiple scalar field models can give rise to non-Gaussian
perturbations of either isocurvature or adiabatic type
\citep{allengrinsteinwise87,kofpog88,SBB89,LM97,Pee99a,Pee99b,Salopek99}.
Alternatively, cosmological defect scenarios (e.g. Vilenkin 1985; Vachaspati
1986; Hill et~al. 1989; Turok 1989; Albrecht \& Stebbins 1992) generally
predict non-Gaussian initial conditions.

The observed abundance of high-redshift cosmic structures contains important
information about the properties of initial conditions on galaxy and clusters
scales.  CMB observations will put constraints on the nature of the initial
conditions (e.g. Pen \& Spergel 1995; Hu et~al. 1997; Verde et~al. 1999; Wang
\& Kamionkowski 1999; Wang et~al. 1999), but will be severely affected by the
presence of noise and foregrounds (e.g. Knox 1999; Tegmark 1998; Bouchet \&
Gispart 1999) and probe rather large scales.

The Gaussian assumption plays a central role in analytical predictions for the
abundance and statistical properties of the first objects to collapse in the
Universe.  In this context, the formalism proposed by Press \& Schechter
\citep{PS74}, with its later extensions and improvements
\citep{PH90,BCEK91,Cole91,LC93} has become the `standard lore' for predicting
the number of collapsed dark matter halos as a function of redshift. However,
even a small deviation from Gaussianity would have a deep impact on those
statistics which probe the tails of the distribution. This is indeed the case
for the abundance of high-redshift objects like galaxies at $z\gap 5$ or
clusters at $z \sim 1$ which correspond to high peaks, i.e. {\it rare events},
in the underlying dark matter density field. Therefore, even small deviations
from Gaussianity might be potentially detectable by looking at the statistics
of high-redshift systems.

The importance of using the mass-function as a tool to distinguish
among different non-Gaussian statistics for the primordial density
field, was first recognized by \citet{LM88}, \citet{Colafrancescoetal89} and,
more recently, by \citet{COS97}, followed by
\citet{RB99}, \citet{RGS99a,RGS99b}, \citet{KST99}, \citet{Willick99},
\citet{AvelinoV99}. To make
predictions on the number counts of high-redshift structures in the
context of non-Gaussian initial conditions, a generalized version of
the PS approach had to be introduced.
 
Such a generalization of the PS formalism has been tested successfully
against N-body simulations \citep{RB99}, but, from a
theoretical point of view, it suffers from the same problems of the
original Press \& Schechter (PS) formulation: it cannot properly
account for the so called {\it cloud-in-cloud} problem, i.e.  the constraint
that bound systems of given mass should not be incorporated in larger
mass condensations of the same catalog\footnote{The so called cloud-in-cloud
problem arises only if one is interested in predicting the local density of
maxima; this feature is not a problem if one focuses the attention to
percolation regions where the ratio of the local density $\rho$ to the
background  density $\rho_b$ is $\rho/\rho_b\sim 200$.}.  Even more important, most of
the PS generalizations to non-Gaussian models proposed in the
literature do not properly take into account the dependence of the
fluctuation field on the smoothing scale.

In this paper we obtain an analytic prediction for the number of dark
matter halos as a function of redshift, within the hierarchical
structure formation paradigm, following the PS formalism.  The
strength of our method is that we are able to properly take into
account the smoothing-scale, or mass, dependence of the probability
distribution function of the primordial density contrast. Obtaining
analytical results in this context is extremely important. Direct
simulations of non-Gaussian fields are generally plagued by the
difficulty of properly accounting for the non-linear way in which
resolution and finite box-size effects, present in any realization of the
underlying Gaussian process, propagate into the statistical properties
of the non-Gaussian field. Moreover, finite volume realizations of
non-Gaussian fields might fail in producing fair samples of the
assumed statistical distribution, i.e. ensemble and (finite-volume)
spatial distributions might sensibly differ (e.g. Zel'dovich 
et~al. 1987). This problem, of course,
becomes exacerbated and hard to keep under control in so far as the
tails of the distribution are concerned. Thus, in looking for the
likelihood of rare events for a non-Gaussian density field, either
exact or approximate analytical estimates should be considered as the
primary tool.

When compared with N-body simulation results the striking feature of the
standard PS algorithm is that it works extremely well in predicting properties
of highly non-linear objects such as the DM haloes, even being based on linear
theory and relatively simple assumptions. This agreement has been proven for
Gaussian initial conditions, and one might wonder whether such a feature also
extends to the non-Gaussian case, where additional non-linear couplings arise
and non-linear gravitational corrections can be important at larger scales,
unlike the Gaussian case.  This extension of the PS prediction will be tested
against N-body simulations in a subsequent paper. For the present application,
however, the primordial non-Gaussianity is assumed to be small and we can
safely assume that departures from the PS prediction are negligible.

The outline of the paper is as follows: in Section 2 we propose a
parameterization of primordial non-Gaussianity that covers a
wide  range of physically motivated models whose non-Gaussianity can be
dialed from zero (the Gaussian limit). We will then relate
analytically the non-Gaussianity parameter to the number of
high-redshift objects.  This step involves the generalization of the
PS formalism to non-Gaussian initial conditions, which in turn
requires an expression for the probability density function of the
non-Gaussian mass-density field as a function of the filtering radius.
In Section 3, we calculate the probability density function for the
density fluctuation field, smoothed on the mass scale typical of
high-redshift objects.  Since the non-Gaussianity is expected to be
small, we can calculate the probability density function by expanding
the `cumulant generator' in powers of the non-Gaussianity parameter
and keeping only linear order terms.  In this context we are able to
obtain an analytic expression for the probability density function of
the smoothed density field.  We generalize the PS approach to
non-Gaussian density fields in Section 4, where we give an analytical
expression for the comoving mass-function of halos formed at a given
redshift, also accounting for the cloud-in-cloud
problem.  Section 5 discusses the association between halos
and high-redshift galaxies. The PS formalism predicts the number of
dark matter halos, but the correspondence to the observed number of
galaxies is not necessarily simple.  Finally in Section 6 we summarize
our main results and draw our conclusions.

\section{Parameterization of primordial non-Gaussianity}

Because of the infinite range of possible non-Gaussian models, we
consider here two models with primordial non-Gaussianity, whose
amplitude can be dialed from zero (the Gaussian limit). Following
\cite{VWHK99} we consider models in which either the density contrast
({\bf model A}) or the gravitational potential ({\bf model B})
contains a part which is the square of a Gaussian random field.  The
physical motivation for this choice is that such a non-Gaussianity may
arise in slow-roll and/or nonstandard (e.g. two-field) inflation
models \citep{Luo94,FRS93,GLMM94,Fanbardeen92}; moreover both models
may be considered as a Taylor expansion of more general non-Gaussian
fields (e.g. Coles \& Barrow 1987; Verde et~al. 1999), 
and are thus fairly generic forms of {\it mild} non-Gaussianity.

We note here that, since in the PS framework the evolution of
perturbations is considered to be linear, this parameterization of
non-Gaussianity is effectively equivalent to non-linear biasing acting
on a truly Gaussian underlying field. The effect of biasing is to
alter the clustering properties of the galaxies with respect to the
dark matter by relating galaxy formation efficiency to the
environment. We do not intend to investigate the clustering properties
of high-redshift objects, we will instead focus on the probability
density function (PDF) of the linear density contrast as a tool to
predict the abundance of dark matter halos as a function of
mass and formation redshift.

\subsection{Linear plus quadratic model for the density or the 
gravitational potential}

As a first step, we want to obtain the PDF for a model where either
the primordial over-density field $\delta\equiv \delta \rho/\rho$, 
or the initial peculiar gravitational potential $\Phi$
is represented by a zero-mean random field $\psi$, given by the
following local transformation ${\cal F}$ on an underlying 
Gaussian field $\phi$:
\begin{equation}
\psi({\bf x}) = {\cal F}[\phi] \equiv \alpha \phi({\bf x}) + 
\epsilon(\phi^2({\bf
x})-\langle\phi^2\rangle) \;,
\label{eq.mapping1}
\end{equation}
with $\alpha$ and $\epsilon$ free parameters of the model.
The homogeneous and isotropic Gaussian process $\phi$ is
assumed to have zero mean, $\langle \phi \rangle=0$, and 
power-spectrum $P_\phi$ to be specified later.
Note that in the limit $\alpha \to 0$, $\psi$ is chi-squared distributed, while
for $\epsilon \to 0$ one recovers the Gaussian case.  As pointed out
before, the model of eq. (\ref{eq.mapping1}) can be thought as
representing the first two
terms of the Taylor expansion of more general non-Gaussian fields
around the Gaussian limit. Cubic or higher order powers of $\phi$ in
such an expansion should be thought as being of order $\epsilon^2$ or
higher.  We will come back to this point in Section 3.3.  Since the
overall normalization of $\psi$ is fixed observationally (e.g. by
measurements of the power-spectrum), effectively there is only one
free parameter in the model.  In what follows we will take either
$\alpha=1$ and $\epsilon$ as the non-Gaussianity parameter or $\alpha=0$
and $\epsilon=1$ to obtain the $\chi^2$ model.

If there was no filtering to take into account or if the transformation 
(\ref{eq.mapping1}) were true for a smoothed field, then the PDF for 
$\psi$ could be simply obtained as follows  
\be
P(\psi)\equiv 
\langle\delta^D(\alpha\phi+\epsilon(\phi^2-\langle\phi^2\rangle) 
- \psi)\rangle
\equiv\int d\phi P(\phi) 
\delta^D(\alpha\phi+\epsilon(\phi^2-\langle\phi^2\rangle) - \psi) 
\label{eq.nosmoothing}
\ee
where $P(\phi)$ is a Gaussian PDF with zero mean and variance 
$\langle \phi^2 \rangle$ and $\delta^D$ is the Dirac delta function. 
The above equation can be also thought as resulting from 
applying the Chapman-Kolmogorov equation:
$P(\psi)=\int d\phi W(\psi|\phi) P(\phi)$ where $ W(\psi|\phi) = 
\delta^D(\psi - {\cal F}[\phi])$ is the transition probability from $\phi$ to
$\psi$ (e.g. Taylor \& Watts 2000). 
The integral over $\phi$ can be performed by writing:
\begin{equation}
P(\psi) = \int d \phi P(\phi) 
\delta^D(\phi - {\cal F}^{-1}[\psi])  {1 \over d{\cal F}/d\phi} \;, 
\end{equation}
which gives:
\begin{equation}
P(\psi) = \left\{2\pi \langle \phi^2 \rangle \left[\alpha^2 + 4
\epsilon \left(\epsilon \langle \phi^2 \rangle +
\psi\right)\right]\right\}^{-1/2} \left[{\cal E}_-(\psi) - 
{\cal E}_+(\psi) \right] \;,
\end{equation}
with
\begin{equation}
{\cal E}_\pm(\psi) \equiv \exp\left\{ \frac{-1}{\epsilon^2\langle
\phi^2 \rangle} \left[ \alpha^2 + 2 \epsilon \left(\epsilon \langle \phi^2
\rangle +\psi \right) \pm \alpha \sqrt{ \alpha^2 + 4 \epsilon 
(\epsilon \langle \phi^2 \rangle + \psi) } \right] \right\} \;.
\end{equation}
 
However, in order to apply the PS formalism, we need to know the PDF
for the linear mass-density field as a function of the smoothing
radius $R$ and redshift $z$. 
To take into account the different evolution of modes inside
the horizon, we assume that the transfer function acts on our 
primordial non-Gaussian field $\psi({\bf x})$, as a convolution. 
We can easily account for the smoothing operation, the effect of the
transfer function and the linear growth of perturbations, by writing
(e.g. Moscardini et~al. 1991): 
\begin{equation}
\delta_R({\bf x},z) = D(z) \delta_R({\bf x}) 
= D(z) \int d^3 y F_R(|{\bf x} - {\bf y}|) 
\psi({\bf y}) \;,
\label{eq.convolution}
\end{equation}
where $D(z)$ is the growing mode of linear perturbations, normalized
to unity at $z=0$, so that $\delta_R({\bf x})$ 
represents the mass density fluctuation field linearly extrapolated
to the present time.   
The isotropic function $F_R(|{\bf x}|)$ can be specified
by its Fourier transform, 
\begin{equation}
{\widetilde F}_R(k) 
\equiv \int d^3 y ~e^{-i \bf k \cdot \bf y} F(|{\bf y}|) =
{\widetilde W}_R(k) T(k) g(k) \;, 
\end{equation}
where $\widetilde{W}_R(k)$ is the Fourier
transform of the assumed low-pass filter (e.g. a spherical top-hat filter),  
$T(k)$ is the transfer function (normalized to unity for $k \to 0$), 
that we will later assume to take the adiabatic CDM form as in 
\citep{BBKS}, with the modifications of \citet{Sugiyama95}.

The function $g(k)$ completes the specification of our model: 
\begin{itemize}
\item {\bf model A}: $g(k)=1$, if we assume that our non-Gaussian 
model applies directly to the primordial density field.
\item {\bf model B}: $g(k) = - \frac{2}{3} (k/H_0)^2\Omega_{0m}^{-1}$,
if the same assumption is made for the gravitational potential 
(its precise form comes from solving the cosmic Poisson equation); 
here $\Omega_{0m}$ denotes the present day density parameter of 
non-relativistic -- baryonic plus dark -- matter.   
\end{itemize} 
As widely discussed by \cite{MMLM91}, 
the overall sign of the
Gaussian to non-Gaussian mapping inherent in these models is a crucial
parameter, which determines most of the non-linear dynamics. 
In our case, we are still free to choose the sign, through the
actual choice of the free parameter $\epsilon$. We finally get
\begin{equation}
\delta_R({\bf x})= \alpha\phi_R({\bf x})+\epsilon\int d^3 y F_R(\mid{\bf
x}-{\bf y}\mid) \phi^2({\bf y}) - C  \;,
\label{eq.deltasmooth}
\end{equation}
where $\phi_R({\bf x})$ is just the smoothed underlying Gaussian field
i.e. the convolution of $\phi$ with $F$, and
\begin{equation}
C \equiv  \epsilon \langle\phi^2\rangle \int d^3 y F_R(\mid{\bf x}-{\bf
y}\mid) \;  
\end{equation}
ensures that $\delta_R$ has zero expectation value. 

\section{PDF of the smoothed density fluctuations}

In this section we derive an approximate analytic expression, valid 
for small values of the non-Gaussian parameter $\epsilon$, for
the PDF as a function of the smoothing radius (i.e. mass scale) and of
the redshift of collapse. This is the key ingredient to develop the
extension of the PS formalism to non-Gaussian fields. The reader
mainly interested
in the application of the PS approach to non-Gaussian density fields
may omit reading this section at a first sitting.

Once the filtering process has been taken into account, 
the density fluctuation PDF can be obtained through a functional or {\it 
path-integral} \footnote{The path-integral approach has been widely applied in the
cosmological context and  in particular to large-scale structure
studies by e.g. \citet{PolitzerWise84,GrinsteinWise86,MLB86,Bert87}.} (e.g. Ramond 1989) 
\begin{eqnarray} 
P(\delta_R) & = & \langle\delta^D\left(\alpha\phi_R({\bf x}) + 
\epsilon\int d^3 y F_R(\mid{\bf x}-{\bf y}\mid)\phi^2({\bf y}) 
- C - \delta_R({\bf x}) \right)\rangle \\ \nonumber 
&=&\int [{\cal D} \phi] {\cal P}[\phi]\int
\frac{d\lambda}{2\pi} \exp\left[ i\lambda \left(\alpha
\phi_R({\bf x}) + \epsilon \int d^3 y F_R(\mid{\bf x}-{\bf
y}\mid)\phi^2({\bf y}) - C - \delta_R({\bf x}) \right)\right] \;,
\label{eq.quaddens-smoothing}
\end{eqnarray}
where, in the second line, we used the integral representation of the Dirac delta function.
The functional integration is over $\phi$ 
configurations in real space weighted by the Gaussian probability
density functional  
\begin{equation}
{\cal P}[\phi] = \exp\left\{-\frac{1}{2}\int d^3 y \int d^3 z
\phi({\bf y}) {\cal K}({\bf y},{\bf z})\phi({\bf z})\right\}
\bigg/ \int [{\cal D} \phi] \exp\left\{-\frac{1}{2}
\int d^3 y \int d^3 z
\phi({\bf y}) {\cal K}({\bf y},{\bf z})\phi({\bf z})\right\} \;,
\label{eq.p-of-phi}
\end{equation}
which has been consistently normalized to unit total probability,
$\int [{\cal D} \phi]{\cal P}[\phi] = 1$. 

Although the previous expressions for the PDF actually gives its form 
at redshift $z=0$, one should keep in mind that the quantity $P(\delta_R) 
d \delta_R$ is redshift-independent, as long as linear evolution
applies. 
   
From now on we will use the following compact notation:
\begin{equation}
\int d^3 y \int d^3 z \phi({\bf y}) {\cal K}({\bf y},{\bf z}) 
\phi({\bf z}) \equiv
(\phi,{\cal K},\phi)
\end{equation}

The symmetric kernel ${\cal K}$ is 
defined as the functional inverse of the two-point correlation function 
$\xi_\phi({\bf y},{\bf w}) = \xi_\phi(|{\bf y} - {\bf w}|)$ of the field 
$\phi$, namely, 
\begin{equation}
\int d^3 y {\cal K}({\bf z},{\bf y}) \xi_{\phi}({\bf y},{\bf w}) = 
\delta^D({\bf z} - {\bf w}) \; .
\label{eq.kei}
\end{equation}

By exploiting the fact that
\begin{equation}
\int d^3 y F_R(\mid {\bf x}-{\bf y} \mid)\phi^2({\bf y}) = 
\int d^3 y \int d^3 z \phi({\bf y})
F_R(\mid {\bf x}-{\bf y} \mid) \delta^D({\bf y}-{\bf z}) \phi({\bf z})
\equiv (\phi,F_R \delta^D ,\phi), 
\label{eq.trick1}
\end{equation}
equation (\ref{eq.quaddens-smoothing}) becomes:
\begin{equation}
P(\delta_R) =\frac{
\int \frac{d\lambda}{2 \pi}e^{- i\lambda \delta_R - i \lambda C} 
\int [{\cal D}\phi] \exp \left\{ - \frac{1}{2} (\phi,{\cal K},\phi) 
+ i\lambda \epsilon(\phi,F_R\delta^D,\phi) + i ({\cal
J}_\lambda,\phi) \right\}} 
{\int [{\cal D}\phi] e^{-\frac{1}{2} (\phi,{\cal K},\phi)}} \;,   
\end{equation}
where we have defined the {\it source} functional 
\begin{equation}
{\cal J}_\lambda ({\bf y}) \equiv \lambda 
\alpha F_R(\mid{\bf x}-{\bf y} \mid)
\end{equation}
and introduced the notation
\begin{equation}
({\cal J}_\lambda,\phi) \equiv 
\int d^3 y {\cal J}_\lambda({\bf y})\phi({\bf y}) \;. 
\end{equation}

The above functional integration can be performed
analytically, by applying the so-called path-integral technique
for {\it composite operators} \citep{cornwalletal74,Hawkingmoss83}. 
Let us briefly sketch the main steps of the procedure.  

We start by defining a new kernel:  
\begin{equation}
{\cal K}^{\prime}_{\lambda}({\bf z},{\bf y}) \equiv {\cal K}({\bf z},{\bf
y}) - i 2\lambda
\epsilon F_R(\mid{\bf y}-{\bf x} \mid)\delta^D({\bf z}-{\bf y}) \;,
\label{eq.keiprime}
\end{equation}
which allows to write 
\begin{equation}
P(\delta_R) = \int\frac{d\lambda}{2 \pi}e^{-i\lambda \delta_R -
i\lambda 
C} \int [{\cal D} \phi] e^{-\frac{1}{2}(\phi,{\cal
K}^{\prime}_{\lambda},\phi) +i({\cal J}_{\lambda},\phi)} \;.
\end{equation}
We then make the change of variable (under which the
path-integral is left unchanged)  
\begin{equation}
\phi({\bf z}) \to \phi({\bf z}) - 
i \int d^3 y \left[{\cal K}^\prime_\lambda\right]^{-1}({\bf z},{\bf y}) 
{\cal J}_{\lambda}({\bf y})
\end{equation}
where $\left[{\cal K}^\prime_\lambda\right]^{-1}$ 
satisfies the integral equation 
\begin{equation}
\int d^3 y {\cal K}^{\prime}({\bf z,y})
\left[{\cal K}^\prime_\lambda\right]^{-1} ({\bf y,w}) = 
\delta^D({\bf z-w}).
\end{equation}
 
We then easily get:
\begin{equation} 
P(\delta_R) = \int\frac{d\lambda}{2 \pi}e^{-i\lambda \delta_R -
i\lambda C - \frac{1}{2}({\cal J}_\lambda,
\left[{\cal K}^\prime_\lambda\right]^{-1},
{\cal J}_\lambda)} \left \{ {\rm Det}[{\cal K}^\prime_\lambda] /
{\rm Det}[{\cal K}] \right \}^{-1/2} \;,
\end{equation}
having used the standard notation for functional determinants:
\begin{equation} 
\{Det[{\cal K}^\prime_\lambda]/Det[{\cal K}]\}^{-1/2} \equiv 
{\int [{\cal D} \phi] \exp[-\frac{1}{2} (\phi,{\cal K}^\prime_\lambda,\phi)] 
\over \int [{\cal D} \phi]\exp[-\frac{1}{2} (\phi,{\cal K},\phi)} = 
\exp\left\{-\frac{1}{2} {\rm Tr} \ln \left[{\bf 1} - i 2\lambda \epsilon (F_R
\delta^D, {\cal K}^{-1})\right] \right\} \;.
\label{eq.jacobian}
\end{equation}
The reader unfamiliar with the functional notation can understand the
last result as a generalization of the identity $\ln [{\rm det}
{\bf M}] = {\rm Tr} [\ln {\bf M}]$, which applies to any symmetric
matrix ${\bf M}$.  
Here ${\bf 1}$ denotes the functional unit matrix, i.e. 
the Dirac delta function; the logarithm of a functional is 
defined as its series expansion and 
\begin{equation}
(F_R\delta^D, {\cal K}^{-1}) \equiv \int d^3 w F_R(|{\bf w} - {\bf x}|) 
\delta^D({\bf y} - {\bf w}) {\cal K}^{-1}({\bf w} - {\bf z}) = 
F_R(|{\bf y} - {\bf x}|) \xi_\phi({\bf y} - {\bf z}) \;.
\end{equation}
Finally, the trace (Tr) of a functional $G({\bf y},{\bf z})$ is defined as 
$\int d^3 y \int d^3 z \delta^D({\bf y} - {\bf z}) G({\bf y},{\bf
z})$. 

Therefore, equation (\ref{eq.jacobian}) becomes:
\begin{equation}
\{Det[{\cal K}^\prime_\lambda]/Det[{\cal K}]\}^{-1/2} 
= \exp\left\{-\frac{1}{2} \!\int \!d^3 y \!\int 
\!d^3 z \delta^D({\bf y} - {\bf z}) \ln \left[ \delta^D({\bf y} - {\bf z}) - 
i 2 \lambda \epsilon F_R(|{\bf y} - {\bf x}|) \xi_\phi({\bf y} -
{\bf z}) \right] \right\}  
\label{eq.jacobian2}
\end{equation}

The PDF for the linearly evolved, smoothed density field takes the 
exact form:
\begin{equation}
P(\delta_R) d\delta_R  = \int\frac{d \lambda}{2 \pi} e^{ - i
\lambda \delta_R + {\cal W}(\lambda)} ~d\delta_R \;,
\end{equation}
where ${\cal W}(\lambda)$ is called the 
{\it cumulant generator}, as 
its series expansion around $\lambda=0$ defines the cumulants (or
irreducible moments) of $\delta_R({\bf x})$. Its exact form is:
\ba
{\cal W(\lambda)} & = &  - i\lambda C - \frac{1}{2} \int d^3 y \int 
d^3 z \biggl\{ \lambda^2 \alpha^2 F_R(|{\bf y}|)
\left[{\cal K}^\prime_\lambda\right]^{-1}({\bf y,z})
F_R(|{\bf z}|)  \nn
& + & \delta^D({\bf y} - {\bf z})
\ln \left[ \delta^D({\bf y} - {\bf z}) - 
i 2 \lambda \epsilon F_R(|{\bf y}|) \xi_\phi({\bf y} -
{\bf z}) \right] \biggr\}  \;.
\label{eq:wfull}
\ea
where the $\lambda$ subscript indicates where the $\lambda$ dependence is
hidden. Note that the ${\bf x}$ dependence has been eliminated by
a mere translation of the origin.
Equation (\ref{eq:wfull}) contains also the exact form for the
generating function for the particular case where the original
non-Gaussian field $\psi$ is chi-squared distributed: this is simply
obtained by setting $\alpha=0$ in the previous expression.   
Note also that the first-order term in the expansion of the
logarithm precisely cancels the $-i\lambda C$ term, so that 
the condition $\langle \delta_R \rangle=0$ is identically satisfied.  

\subsection{Kernel inversion} 

In order to solve the remaining
integrals we need to find an expression for the functional $[{\cal
K}^{\prime}_\lambda]^{-1}$.  Let us start by finding an expression for
the kernel ${\cal K}$, given that its inverse is just the auto-correlation
function of the Gaussian field $\phi$, namely,
\begin{equation}
\int d^3 y {\cal K}({\bf w} - {\bf y})\xi_\phi(\mid {\bf y}-{\bf
z}\mid)=\delta^D({\bf w}-{\bf z}) \;. 
\label{eq.defkei}
\end{equation} 

Fourier transforming both sides of eq. (\ref{eq.defkei}) we easily find \citep{PolitzerWise84,Bert87}
\begin{equation}
\widetilde{{\cal K}}(k)=1/P_{\phi}(k)
\end{equation}
where $P_\phi(k)$ is the power-spectrum of the underlying Gaussian
field $\phi$. 
Therefore \newline
\mbox{${\cal K}({\bf w},{\bf y})\equiv {\cal K}(\mid {\bf w}-{\bf
y}\mid)\equiv{\cal K}(r)$} reads
\begin{equation}
{\cal K}(r)= {1 \over (2 \pi)^3} \int d^3 k \exp(i
{\bf k} \cdot {\bf r})\frac{1}{P_{\phi}(k)} = {1 \over 2 \pi^2} 
\int_0^\infty dk k^2 j_0(kr) \frac{1}{P_{\phi}(k)} \;. 
\end{equation}

To find an expression for  $\left[{\cal K}^{\prime}_\lambda\right]^{-1}$ 
one would like to proceed in an analogous
way. However, owing to the absence of translational invariance of 
${\cal K}^\prime$, going to momentum space does not help. 
The technique we are then going to use is to expand 
$\left[{\cal K}^{\prime}_\lambda\right]^{-1}$ in
powers of the non-Gaussianity parameter $\epsilon$, as follows:
\begin{equation}
\left[{\cal K}^\prime_\lambda\right]^{-1} ({\bf y},{\bf z}) =  
\sum_{n=0}^\infty (2 i \epsilon \lambda)^n {\cal R}^{(n)}({\bf y},{\bf z}) 
\end{equation}
where the coefficients ${\cal R}^{(n)}$ are obtained recursively from 
\begin{equation}
{\cal R}^{(n)}({\bf y},{\bf z}) 
= \int d^3 w \xi_\phi(|{\bf y} - {\bf w}|) F_R(|{\bf w}|) 
{\cal R}^{(n-1)}({\bf w},{\bf z})  \ \ \ \ \ (n>0) \;, 
\end{equation}
with ${\cal R}^{(0)}({\bf y},{\bf z}) = \xi_\phi(|{\bf y}
- {\bf z}|)$.
Similarly, in Fourier space we have
\begin{equation}
{\widetilde {\cal R}}^{(n)}({\bf k},{\bf k}^\prime) = 
P_\phi(k) \int {d^3 q \over (2 \pi)^3} 
{\widetilde F}_R( q) {\widetilde {\cal R}}^{(n-1)}
({\bf q} + {\bf k}, {\bf k}^\prime) \;,
\end{equation} 
with ${\widetilde {\cal R}}^{(0)}({\bf k},{\bf k}^\prime) = 
( 2\pi)^3 P_\phi(k) 
\delta^D({\bf k} + {\bf k}^\prime)$. 

From equation (33) we find:
\begin{equation}
{\widetilde {\cal R}}^{(1)}({\bf k},{\bf k}^\prime) = 
P_\phi(k) P_\phi(k^\prime) {\widetilde F}_R(|{\bf k} + {\bf k}^\prime|) 
\end{equation}
and, for $n\geq 2$, 
\begin{eqnarray}
{\widetilde {\cal R}}^{(n)}({\bf k},{\bf k}^\prime) & = &
P_\phi(k) P_\phi(k^\prime)
\int \frac{d^3 q_1}{(2\pi)^3} \cdots \int \frac{d^3 q_{n-1}}{(2\pi)^3}
{\widetilde F}_R(q_1) \cdots {\widetilde F}_R(q_{n-1})
\\ \nonumber
& \times & 
{\widetilde F}_R(|{\bf q}_1 + \cdots + {\bf q}_{n-1} + 
{\bf k} + {\bf k}^\prime|)P_\phi(|{\bf q}_1 + {\bf k}|) 
\cdots P_\phi(|{\bf q}_1 + \cdots + {\bf q}_{n-1} + {\bf k}|) \;.
\end{eqnarray}
Note that, while in principle the expressions for ${\cal K}^\prime$, 
${\cal R}^{(n)}$ and $\widetilde{ {\cal
R}}^{(n)}$ should be symmetrized, for the purpose of calculating 
the PDF or the cumulants, this operation is not needed. 

\subsection{Cumulant generator}

At this point we are ready to provide an expression for the cumulant 
generator, by first expanding it in powers of $\lambda$. 
We have:  
\begin{equation}
{\cal W}(\lambda) \equiv \sum_{n=2}^\infty \frac{(i\lambda)^n}{n!}
\mu_{n,R} \;, 
\label{eq:wlinear}
\end{equation} 
where $\mu_{n,R}$ denotes the cumulant of order $n$ of the smoothed 
density contrast $\delta_R$.

The variance, the skewness and the
kurtosis of the smoothed non-Gaussian density field are,
respectively \footnote{The integral over k in the sub-leading term of the
variance can diverge for certain choices of the power-spectrum 
of the underlying Gaussian
field and transfer function. In these cases one should bear in mind
that any physical process originating the underlying field will
necessarily provide the $\phi$ power-spectrum with both 
infrared and ultraviolet cutoffs (the present-day horizon and the
reheating scale, respectively, in the case of inflation-generated 
perturbations). In case this contribution dominates
over the leading-order contribution even for $\epsilon \ll 1$ one can
postulate the existence of a term proportional to $\epsilon^2$ in
eq. (1) that will cancel out the sub-leading term in the variance. 
Alternatively eq. (1) and eq. (\ref{eq.variancefull}) can be
renormalized by choosing $\alpha$ and
$A_{\phi}$ such that
$\mu_{2,R}\simeq\mu_{2,R}^{(1)}$ when $\epsilon \ll 1$. Our calculations
still apply in this case provided one interprets $\epsilon$ as 
$\epsilon/\alpha^2$.}, 
\ba
\mu_{2,R} & \equiv & \sigma^2_R \equiv \langle \delta_R^2 \rangle 
= {\alpha^2 \over
2 \pi^2} \int_0^\infty dk k^2 {\widetilde F}^2_R(k) P_\phi(k) \nn
& +&
\frac{\epsilon^2}{2 \pi^4} \int_0^\infty d k k^2 P_\phi(k)
\int_0^\infty d k^{\prime} k^{\prime^2} P_\phi(k^\prime) 
\int_0^1 d \mu ~{\widetilde F}_R^2(\sqrt{k^2 + k^{\prime^2} +2 k
k^{\prime} \mu}) \;,
\label{eq.variancefull}
\ea

\begin{eqnarray}
\mu_{3,R} & \equiv & \langle \delta_R^3 \rangle 
= {3 \epsilon \alpha^2 \over 2 \pi^4} \int_0^\infty\! d k k^2 
{\widetilde F}_R(k) P_\phi(k)
\int_0^\infty \! d k^{\prime} k^{\prime^2} {\widetilde F}_R(k^\prime) 
P_\phi(k^\prime) 
\int_0^1\! d \mu ~{\widetilde F}_R(\sqrt{k^2\! +\! k^{\prime^2}\! +\! 
2 k k^{\prime}\! \mu})
\\ \nonumber
& + & 8 \epsilon^3 \int \frac{d^3 k_1}{(2\pi)^3} 
\int \frac{d^3 k_2}{(2\pi)^3} \int \frac{d^3 k_3}{(2\pi)^3} 
P_\phi(k_1) P_\phi(k_2) P_\phi(k_3) {\widetilde F}_R(|{\bf k}_1 - {\bf
k}_2|) {\widetilde F}_R(|{\bf k}_2 - {\bf k}_3|)
{\widetilde F}_R(|{\bf k}_3 - {\bf k}_1|) \;. 
\end{eqnarray}
and

\begin{eqnarray}
\mu_{4,R} & \equiv &\!\!\! \langle \delta_R^4 \rangle - 3 \langle \delta_R^2
\rangle^2 \nn
& =&  48 \epsilon^2 \alpha^2 \! \int \! \frac{d^3 k_1}{(2 \pi)^3} \! 
\int \! \frac{d^3 k_2}{(2 \pi)^3}\!  \int \! \frac{d^3 k_3}{(2 \pi)^3}\! 
P_\phi(k_1) P_\phi(k_2) P_\phi(|{\bf k}_3 + {\bf k}_1|) 
{\widetilde F}_R(k_1) {\widetilde F}_R(k_2) 
{\widetilde F}_R(k_3) 
\\ \nonumber
& \times & 
{\widetilde F}_R(|{\bf k}_1 + {\bf k}_2 + {\bf k}_3|) + 
96 \epsilon^4 \int \frac{d^3 k_1}{(2 \pi)^3}
\int \frac{d^3 k_2}{(2 \pi)^3} \int \frac{d^3 k_3}{(2 \pi)^3}
\int \frac{d^3 k_4}{(2 \pi)^3} P_\phi(k_1) P_\phi(k_2)  P_\phi(k_3) 
P_\phi(k_4) 
\\ \nonumber
& \times & 
{\widetilde F}_R(|{\bf k}_1 - {\bf k}_2|)
{\widetilde F}_R(|{\bf k}_2 - {\bf k}_3|) {\widetilde F}_R(|{\bf k}_3
- {\bf k}_4|) {\widetilde F}_R(|{\bf k}_4 - {\bf k}_1|) \;. 
\end{eqnarray}

More in general, the cumulants are made of two 
contributions: a leading term of order $\epsilon^{n-2}$ plus a 
subleading term of order $\epsilon^n$,  
\begin{equation}
\mu_{n,R} \equiv \epsilon^{n-2} \mu_{n,R}^{(1)}
+ \epsilon^n \mu_{n,R}^{(2)} \;,
\end{equation}
where 
\begin{equation}
\mu_{n,R}^{(1)}\! =\! 2^{n-3} n! ~\alpha^2 \int \frac{d^3 k}{(2\pi)^3} \int 
\frac{d^3 k^\prime}{(2\pi)^3} {\widetilde F}_R(k_1) {\widetilde F}_R(k_2)
{\widetilde {\cal R}}^{(n-2)}({\bf k},{\bf k}^\prime) 
\end{equation}
and 
\begin{equation}
\mu_{n,R}^{(2)} = 2^{n-1} (n-1)!\!\! 
\int \!\frac{d^3 k_1}{(2\pi)^3} \cdots\! \int \! \frac{d^3 k_n}{(2\pi)^3}
P_\phi(k_1) \cdots P_\phi(k_n) {\widetilde F}_R(|{\bf k}_1 - {\bf
k}_2|) \cdots  {\widetilde F}_R(|{\bf k}_{n-1} - {\bf k}_n|)
{\widetilde F}_R(|{\bf k}_n - {\bf k}_1|) \;. 
\end{equation}

Note that, as anticipated, the sign of the parameter $\epsilon$, for a 
given model, fully determines the sign of the skewness  
as well as of all the other odd-order cumulants. 

The cumulants for the filtered chi-squared model are immediately
recovered by taking $\alpha=0$ and $\epsilon=1$ in the previous
expressions, so that $\mu_{n,R}=\mu^{(2)}_{n,R}$.  It is worth
noticing that eq. (42) supplies {\it all} the cumulants of a chi-squared
field smoothed on scale $R$. This is an interesting result for models
such as that recently proposed by \citet{Pee99a,Pee99b}, where
non-Gaussian isocurvature fluctuations are obtained with a chi-squared
distributed density field.

\subsubsection{Linear order in $\epsilon$}

\begin{figure}
\begin{center}
\setlength{\unitlength}{1mm}
\begin{picture}(140,55)
\includegraphics{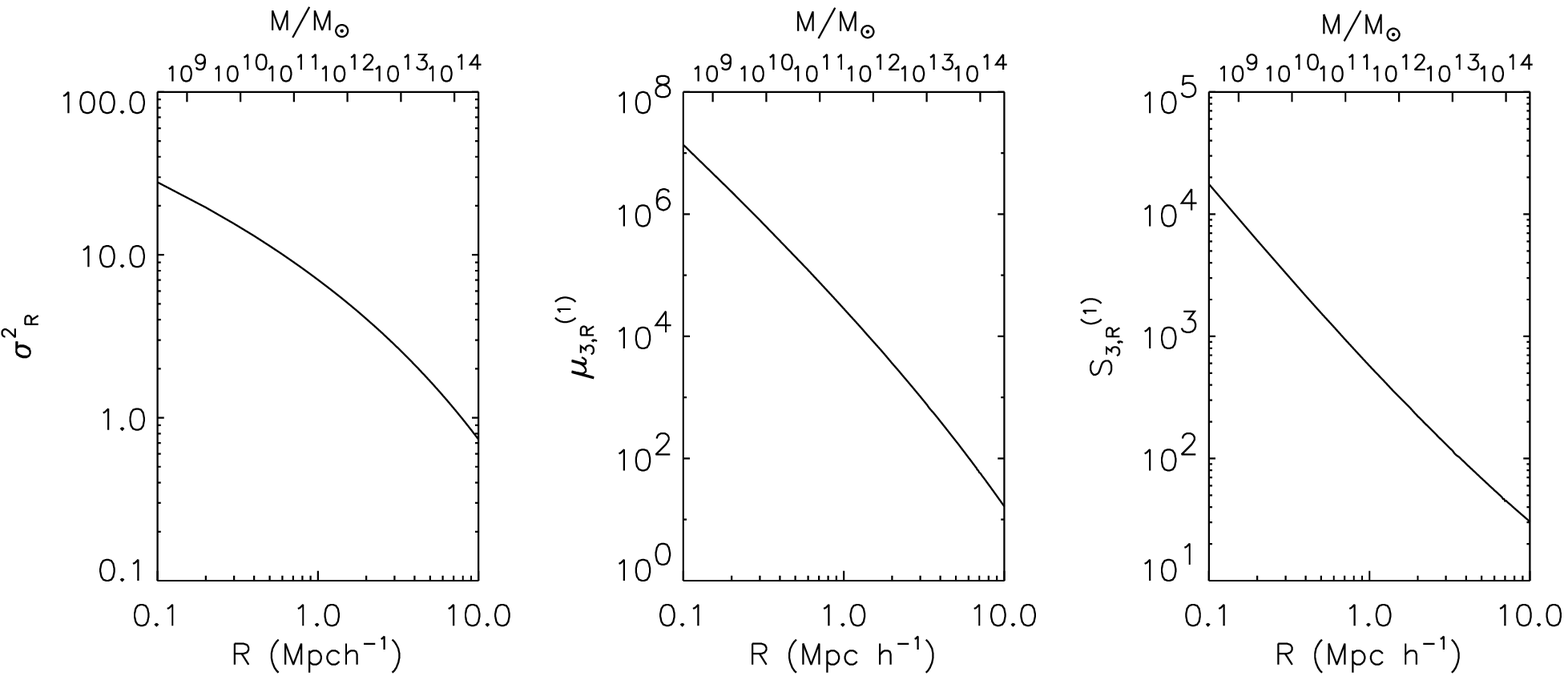}
\end{picture}
\end{center}
\caption{}
\label{fig:sigmamuq}
\end{figure}

To first order, the skewness depends linearly on
$\epsilon$, therefore let us define
\be
S_{3,R}\equiv\epsilon S_{3,R}^{(1)}=\epsilon\mu_{3,R}^{(1)}/(\mu_{2,R}^{(1)})^2,
\ee
where 
\be
\mu_{2,R}^{(1)}={\alpha^2 \over
2 \pi^2} \int_0^\infty dk k^2 {\widetilde F}^2_R(k) P_\phi(k)
\ee
and
\be
\mu_{3,R}^{(1)}=\langle \delta_R^3 \rangle 
= {3 \alpha^2 \over 2 \pi^4} \int_0^\infty d k k^2 
{\widetilde F}_R(k) P_\phi(k)
\int_0^\infty d k^{\prime} k^{\prime^2} {\widetilde F}_R(k^\prime) 
P_\phi(k^\prime) 
\int_0^1 d \mu ~{\widetilde F}_R(\sqrt{k^2 + k^{\prime^2} + 
2 k k^{\prime} \mu})
\ee
To this order in $\epsilon$ 
there is no contribution from higher order moments. 
On very large scales,
where the transfer function becomes unity, and for a power-law 
$\phi$ power-spectrum ($P_\phi=A_\phi k^n$ or $P_\phi=A_\phi
k^{n-3}$ for model A or B, respectively), this leading-order skewness
parameter $S_{3,R}$ becomes scale-independent,  
thus mimicking the behaviour induced by the gravitational
instability \citep{Pee80}.
For model A the variance, the skewness
$\mu_{3,R}^{(1)}$ and  skewness parameter
$S_{3,R}$ are plotted vs. the radius $R$ and the mass $M$ 
in solar masses, $M\propto R^3$, 
in Figure \ref{fig:sigmamuq} with the following assumptions:
$\alpha=1$, top-hat filter in real space, transfer function as in
\citep{Sugiyama95}, with baryonic matter density parameter $\Omega_{0b}=0.015 
h^{-2}$, $h=0.65$,
$\Omega_{0m}=0.3$, and scale-invariant primordial power-spectrum 
$P_\phi \propto k$, normalized so that the present day
r.m.s. fluctuation on a sphere of $8~h^{-1}$ Mpc is  
$\sigma_8=0.99$ \citep{Vianaliddle98}, which simultaneously allows 
to best fit the local cluster abundance and the {\it COBE} data
(e.g. Tegmark 1996). 
In what follows we will always assume that the approximation 
$\sigma^2_R\equiv \mu_{2,R}^{(1)}$ applies on all considered scales.  
As detailed below, at $z=0$, for $\epsilon \leq 0.01$ in model A we
will recover the
standard PS mass-function on clusters scales, so that the relation
between the observed abundance of clusters and the value of $\sigma_8$
keeps unchanged, in spite of our non-Gaussian assumption. 

\begin{figure}
\begin{center}
\setlength{\unitlength}{1mm}
\begin{picture}(140,55)
\includegraphics{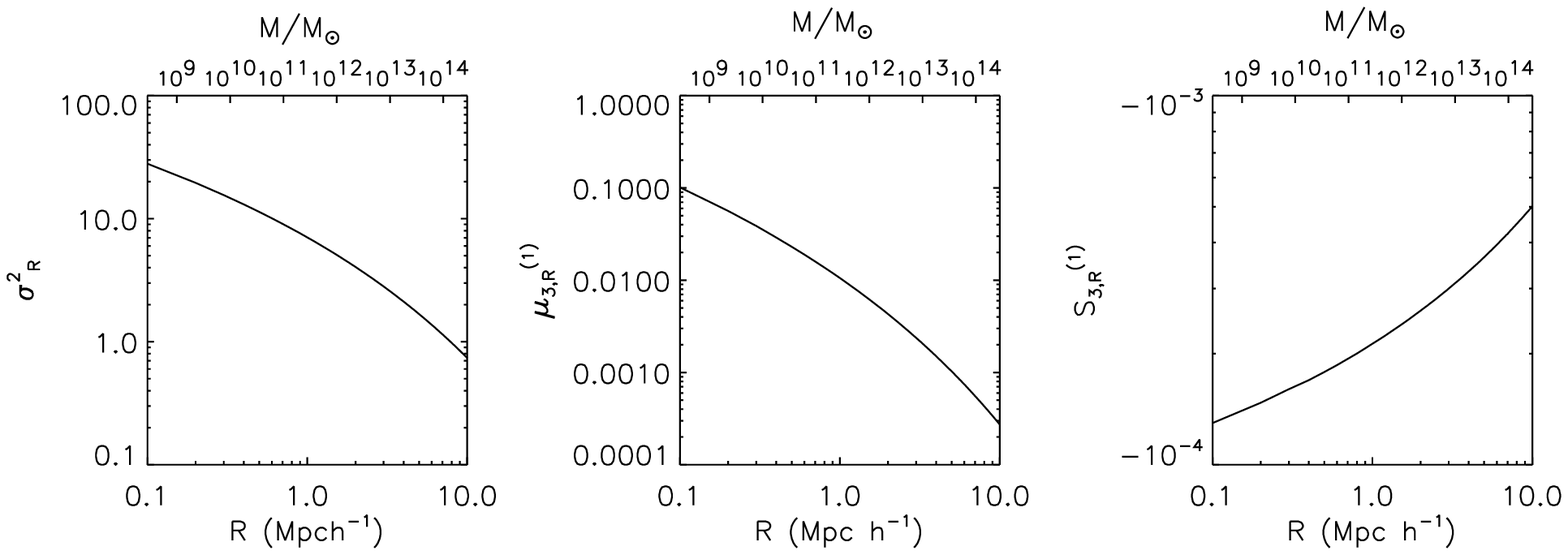}
\end{picture}
\end{center}
\caption{}
\label{fig:sigmamuqpot}
\end{figure}

To deal with model B we take $P_\phi \propto k^{-3}$ and, as
above, we normalize the mass-density variance as in model A, 
with the same choice of cosmological parameters and transfer function. 
The variance, the skewness $\mu_{3,R}^{(1)}$ and  skewness parameter
$S^{(1)}_{3,R}$ are plotted vs. the radius $R$ and the mass $M$ 
in solar masses, $M\propto R^3$, 
in Figure \ref{fig:sigmamuqpot} with the same assumptions as for model A in
Figure \ref{fig:sigmamuq}. The variance for model B to linear order in
$\epsilon$ is identical to the variance for model A, but the skewness and
skewness parameter are different: in particular notice that $S_{3,R}^{(1)}$
for model B has the opposite sign of $\epsilon$ and, for a given
value of $\epsilon$, its amplitude is many orders 
of magnitude smaller than for 
model A. It is important to realize, that both signs for $\epsilon$ 
are generally allowed. In the inflationary case, both the sign and the
magnitude of $\epsilon$ are related to the inflationary slow-roll
parameters $\epsilon_{\rm infl}$ and $\eta_{\rm infl}$ 
\citep{GLMM94,Gangui94,Wang99,Ganguimartin99}. 
In this model, the PS mass-function at $z=0$ is practically recovered
on all scales, for all $|\epsilon|~\lap 200$. 



\subsubsection{Quadratic order in $\epsilon$}
For small deviations from Gaussianity, a first-order expansion in
$\epsilon$ is a valid approximation. 
We give here the expressions for the relevant quantities also to second order,
but in any application we will neglect second or higher-order corrections. 

When expanding the cumulant generator to second order in 
$\epsilon$ we obtain
that the variance has a contribution $\propto \epsilon^2$
as in eq. (\ref{eq.variancefull}), the skewness remains the same as to
first-order in $\epsilon$, the next to leading term being $\propto \epsilon^3$,
but there is a non-vanishing contribution to the kurtosis $\propto
\epsilon^2$, namely 
\ba
\mu_{4,R}& = &\epsilon^2\mu_{4,R}^{(1)}= 48 \epsilon^2\alpha^2\int\! 
\frac{d^3 k_1}{(2 \pi)^3} 
\int\! \frac{d^3 k_2}{(2 \pi)^3}  \int \! \frac{d^3 k_3}{(2 \pi)^3} 
P_\phi(k_1) P_\phi(k_2) P_\phi(|{\bf k}_3 + {\bf k}_1|) \nn
&\times &  {\widetilde F}_R(k_1) {\widetilde F}_R(k_2) 
{\widetilde F}_R(k_3)  
{\widetilde F}_R(|{\bf k}_1 + {\bf k}_2 + {\bf k}_3|)
\ea

For both models, the leading-order kurtosis parameter
$S_{4,R}=\epsilon^2\mu_{4,R}^{(1)}/(\mu_{2,R}^{(1)})^3$ becomes 
scale-independent if the power-spectrum is a power-law; 
therefore on large scales, where the transfer function is unity, 
$S_{4,R}$ becomes scale-independent.

\subsection{Higher order non-Gaussian contributions}

What happens to our results if we allow for higher order terms 
in the original definition of our primordial non-Gaussian field
$\psi$? Let us modify, for instance, our definition by adding a cubic
term, as follows, 
\begin{equation}
\psi({\bf x}) = \alpha \phi({\bf x}) + \epsilon(\phi^2({\bf x}) -
\langle \phi^2\rangle) + \epsilon^2 \beta \phi^3({\bf x}) \;, 
\end{equation} 
where $\beta$ is a new independent parameter, which we assume to
be of order unity. Such a cubic term might be easily accounted for in the 
functional integral approach, by using the so-called `integration by
parts' relation (e.g. Ramond 1989). 
For the purpose of the present paper, however, we can account for such
a cubic term through the modifications it induces in the 
lowest-order cumulants. The sub-leading term of the variance 
would in fact be modified by the addition of  
\begin{equation} 
\epsilon^2 \Delta \mu^{(2)}_{2,R} = 6 \epsilon^2 \frac{\beta}{\alpha}
\langle\phi^2 ~\rangle \mu_{2,R}^{(1)} \;,  
\end{equation}
which therefore appears as a `renormalization' of the 
leading-order variance. 
The sub-leading skewness would also get an extra 
contribution of the same order $\epsilon^3$, which we will not write here. 
The kurtosis, instead, would be modified already to
leading order, gaining the extra piece
\begin{equation}
\epsilon^2 \Delta \mu^{(1)}_{4,R} = {1\over 2} \epsilon^2 \alpha \beta
\mu_{4,R}^{(1)} \;,   
\end{equation}
which still has the nice feature of appearing as a renormalization of
the previously calculated leading-order kurtosis. 
 
At this point we can derive a general \footnote{The key
assumption that allowed us to obtain such a general treatment of a mildly
non-Gaussian field is that it can be expanded as a local functional of
an underlying Gaussian field. On the
other hand, any form of non-locality such that it can be 
expressed as a convolution in real space can also be handled in a similar way, 
by suitably modifying our definition of the function
$F_R({\bf x})$.} and self-consistent
approximation to the cumulant generator (and hence for both the
differential and cumulative probability) up to order $\epsilon^2$:
\begin{equation}
{\cal W}(\lambda) = - {\lambda^2 \over 2} \mu_{2,R}^{(1)} 
- \epsilon {i \lambda^3 \over 6} \mu^{(1)}_{3,R} + 
\epsilon^2 \left[ - {\lambda^2\over 2} (\mu^{(2)}_{2,R}+6 \frac{\beta}{\alpha}\langle \phi^2 \rangle \mu_{2,R}^{(1)} ) 
+ {\lambda^4 \over 24} \left(1 + \frac{1}{2} \alpha \beta \right) 
\mu^{(1)}_{4,R} \right] \;. 
\label{eq:approxgen}
\end{equation}
This is the explicit expression for ${\cal W}(\lambda)$; no further
terms could
appear from higher-order non-Gaussianity to this order in $\epsilon$.
Substituting this expansion in eq. (26) one gets an approximate
expression for the PDF as an integral over $\lambda$, which is valid
for suitably small values of the non-Gaussian parameter $\epsilon$. 

For the rest of our calculations, however, we will assume that
departures from Gaussianity are small and therefore we will retain
only linear-order terms.

\section{Press-Schechter approach to non-Gaussian density fields}

To obtain the abundance of dark matter halos as a function of
filtering radius $R$ (or mass $M\propto R^3$) and redshift of collapse
$z_c$, one should first obtain the conditional probability that the
density contrast equals the threshold for
collapse $\delta_c$, when filtered on scale $R$, provided it is below
it on any larger scale $R^\prime$. 
In the Gaussian case, and for sharp-k-space 
filter, this problem has been solved by a number of authors
\citep{PH90,Cole91,BCEK91} by rephrasing it in terms of the problem
of barrier first-crossing by a Markovian random walk.  In the
non-Gaussian case and/or for other types of filters, different
techniques might also be useful.  In particular, an alternative formulation,
originally proposed by \citet{Jedamzik95} and successively implemented
by \citet{Yanoetal96}, \citet{NagashimaGouda97} and \citet{LeeShandarin98}, allows to
reduce the problem to the solution of the integral equation, 
\be
P(>\delta_c|z_c,M) = \frac{1}{{\bar \rho}_{0m}} \int_0^\infty d
M^\prime P(M|M^\prime) M^\prime n(M^{\prime},z_c) \;, 
\ee

where ${\bar  \rho}_{0m}$ is the present-day mean density of non-relativistic
matter, $P(>\delta_c|z_c,M)$ is the probability that $\delta_M$ lies
above the threshold $\delta_c$ (i.e. the fraction of volume where this
happens) at a given redshift $z_c$ and $n(M,z_c)$ is the required
comoving mass-function for halos of mass between $M$ and $M+d M$ which
formed at $z_c$. The function $P(M|M^\prime)$ denotes the conditional
probability of finding a region with mass $M$ overdense by $\delta_c$
or more, given that it is included in an {\it isolated} region of mass
$M^\prime$ $(>M)$.

Let us start by first finding an expression for the L.H.S. of this
equation, i.e. for the level-crossing probability
$P(>\delta_c|z_c,M)$.
  
\begin{figure}
\begin{center}
\setlength{\unitlength}{1mm}
\begin{picture}(140,55)
\includegraphics{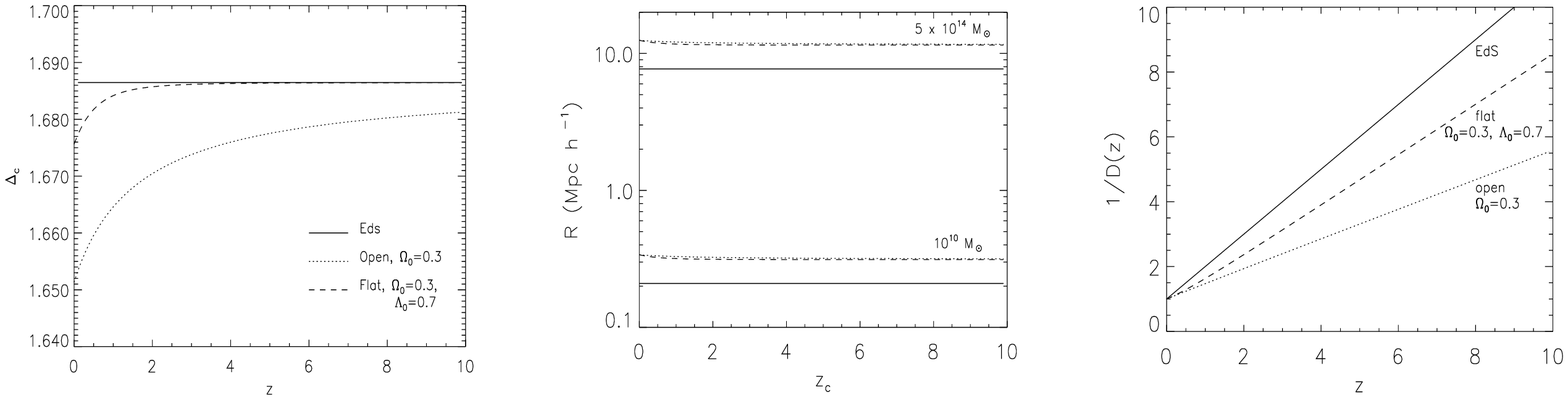}
\end{picture}
\end{center}
\caption{}
\label{fig:redshiftdep}
\end{figure}

In looking at this problem, it is convenient to think of the density
fluctuation as being time-independent while giving a redshift
dependence to the collapse threshold $\delta_c(z_c) \equiv
\Delta_c(z_c)/D(z_c\mid\Omega_{0m},\Omega_{0\Lambda} )$; $\Delta_c$ is
the linear extrapolation of the over-density for spherical collapse:
it is $1.686$ in the Einstein-de Sitter case, while it slightly
depends on redshift (see Figure \ref{fig:redshiftdep}) for more
general cosmologies (e.g. Kitayama \& Suto 1996); $\Omega_{0\Lambda}$
denotes the closure density of vacuum energy today.  Here $D(z\mid
\Omega_{0m},\Omega_{0\Lambda})$ denotes the general expression for the
linear growth factor, which depends on the background cosmology: the
redshift dependence of $D(z_c\mid\Omega_{0m},\Omega_{0\Lambda})^{-1}$
is shown in Figure \ref{fig:redshiftdep} for three different
cosmological models.

Using the spherical isothermal collapse model \citep{GG72}, it is possible to
relate the mass $M$ of a dark halo to the Lagrangian (pre-collapse) comoving
length $R$ (the sphere of radius $R$ will give rise to an object that 
contains the mass M within the virialization radius) (e.g. White,
Efstathiou \& Frenk 1993), 
\begin{equation}
R=\frac{2^{1/2}[V_c/100{\rm km s}^{-1}]}{\Omega_{0m}^{1/2}
(1+z_c)^{1/2} f_c^{1/6}}
\label{eq:massradius}
\end{equation}
where $R$ is in units of Mpc h$^{-1}$. Here  $V_c$ is the circular 
velocity required
for centrifugal support in the potential of the halo, given by 
$M=V_c^3/(10GH(z_c))$,
where $H$ denotes the Hubble constant,
(e.g. Heavens \& Jimenez 1999), and $f_c$ is the density contrast at 
virialization of the
newly-collapsed object relative to the background. This is adequately
approximated by $f_c=178/\Omega_m^{0.6} (z_c)$ (e.g. Eke, Cole \&
Frenk 1996).


In Figure \ref{fig:redshiftdep} the dependence of
(\ref{eq:massradius}) on the cosmology and the redshift of collapse 
is shown. It is clear that this dependence is very weak and can be
ignored for our purposes.
For a chosen power-spectrum shape and normalization, the dependence on the
cosmological model of the PDF and the level excursion probability is
therefore confined in $\delta_c(z_c)$. In other words, changing the
cosmology is essentially equivalent to changing $z_c$ according to the
right panel of Figure \ref{fig:redshiftdep}.  

Following the PS formalism, we can write
\begin{equation}
P(>\delta_c|z_c,R) = \int_{\delta_c(z_c)}^\infty d \delta_R 
P(\delta_R) = \int_{\delta_c(z_c)}^\infty d \delta_R 
\int_{-\infty}^\infty {d \lambda  \over 2 \pi}
e^{-i \lambda \delta_R + {\cal W}(\lambda)} \;,
\end{equation}
which, exchanging the order of integrations and integrating over
$\delta_R$, yields the exact and general expression
\begin{equation} 
P(>\delta_c|z_c,R) = {1 \over 2 \pi i} 
\int_{-\infty}^\infty {d \lambda  \over \lambda} 
\exp\left[-i \lambda \delta_c(z_c) + {\cal W}(\lambda) \right] 
+ {1 \over 2} \;.
\label{eq.exactandgeneral}
\end{equation}

This is the exact expression for the probability that at redshift
$z_c$, the fluctuation on scale $R$
exceeds some critical value $\delta_c$. From this equation 
it is possible to obtain an approximate expression for the cumulative 
probability $P(>\delta_c|z_c,R)$ by expanding 
$W(\lambda)$ of eq. (\ref{eq:wfull}) to a given order in $\epsilon$.
On the other hand, from knowledge of the cumulants of the  
cosmological density field smoothed on scale $R$, up
to some order,  it is possible to obtain an
approximate expression
for the level excursion probability $P(>\delta|R)$.

The integral (\ref{eq.exactandgeneral}) can be solved analytically by
using the {\it saddle-point} technique
as in \citep{Fry86} and \citep{LM88}. We first perform a {\it Wick
rotation}, $\lambda \to i\lambda$, in the complex $\lambda$ plane,
to get rid of the oscillatory behaviour of the integrand. 
This gives 
\begin{equation} 
P(>\delta_c|z_c,R) = {1 \over 2 \pi i} 
\int_{-i\infty}^{i\infty} {d \lambda  \over \lambda} 
\exp\left[- \lambda\delta_c(z_c) + {\cal M}(\lambda) \right] + {1 \over 2} \;,
\end{equation}
with ${\cal M}(\lambda) \equiv {\cal W}(-i\lambda)$. 
Next, let us introduce the {\it effective action} $G(\delta_{\rm eff})$,
defined as the Legendre transform of ${\cal M}$, namely 
$G(\delta_{\rm eff}) = \lambda \delta_{\rm eff} - {\cal M}(\lambda)$,
with $\delta_{\rm eff} = d {\cal M}/d \lambda$. 
From the definition of $\cal M$ one has
\begin{equation}
P(>\delta_c|z_c,R) = {1 \over 2 \pi i} 
\int_{G^\prime=-i\infty}^{G^\prime=i\infty} d \delta_{\rm eff} 
{G^{\prime \prime}(\delta_{\rm eff}) \over 
G^\prime(\delta_{\rm eff})} 
\exp\left[\left(\delta_{\rm eff} - \delta_c(z_c)\right) 
G^\prime(\delta_{\rm eff}) - G(\delta_{\rm eff})\right] 
+ {1 \over 2} \;. 
\label{eq:exacteffaction}
\end{equation}
For large thresholds, $\delta_c(z_c) \gg 1$, the above integral is
dominated by stationary points of the exponential. These occur at 
$G^{\prime \prime}(\delta_{\rm eff})(\delta_c(z_c) -
\delta_{\rm eff}) =0$, i.e. at $\delta_{\rm eff}= \delta_c(z_c)$,
since $G^{\prime \prime} (\delta_{\rm eff}) = (d^2 {\cal M}/d
\lambda^2)^{-1}>0$ [as ${\cal M}$ must be a convex function of its argument
(e.g. Fry 1985)]. 
A saddle-point evaluation of this integral then gives \citep{LM88}
\begin{equation}
P(>\delta_c|z_c,R) \approx {1 \over \sqrt{2 \pi}} {(G^{\prime
\prime})^{1/2} 
\over G^\prime} \exp(-G)\bigg|_{\delta_{\rm eff}=\delta_c(z_c)} \;.
\label{eq:pdeltacsaddlepoint}
\end{equation}
Note that the remaining $1/2$ term on the r.h.s. of equation 
(\ref{eq:exacteffaction}) has been
exactly canceled by the pole residual of the integral at 
$G^\prime(\delta_{\rm eff})= 0$. 
The latter formula provides an accurate approximation of the 
level-crossing probability $P(>\delta_c)$, provided the cumulant generator
${\cal M}(\delta_{\rm eff})$ is analytic for all finite values of its
argument, which is indeed the case for our approximated
expression in eq. (\ref{eq:approxgen}), but certainly not
for the exact form of eq. (\ref{eq:wfull}). The validity of our last result
will then rely on the smallness of the non-Gaussianity parameter $\epsilon$. 
It is important to notice here that since - as it is clear from
Figure \ref{fig:sigmamuq} - the skewness parameter is not 
small everywhere, we expect the PDF to be very sensitive to small $\epsilon$
at least for model A.
 
The approximate form of the effective action resulting from
Legendre transforming eq. (\ref{eq:approxgen}) is
\begin{equation}
G(\delta_{\rm eff}) \approx {\delta_{\rm eff}^2 \over 2 \sigma_R^2} 
\left\{1 - \epsilon^2\left(6  \frac{\beta}{\alpha} \langle\phi^2\rangle + 
\frac{\mu_{2,R}^{(2)}}{\sigma_R^2} \right) - 
\frac{S_{3,R}}{3} \delta_{\rm eff} +
\left[\frac{1}{4} S_{3,R}^2 -
\frac{1}{12} \left(1+ \frac{1}{2}\alpha\beta\right) S_{4,R}\right] 
\delta_{\rm eff}^2 \right\} \;.
\end{equation}

The level crossing probability $P(>\delta_c)$ to second order in
$\epsilon$ can be obtained from (\ref{eq:pdeltacsaddlepoint}), 
with (57) and the following substitution:
\be
\frac{(G'')^{1/2}}{G'}=\frac{\sigma_R}{\delta_{\rm eff}} \left[ 1 +
\frac{\epsilon^2}{24\alpha }\left(72 \beta \langle\phi^2 \rangle
\sigma^2_R-\alpha^2\beta\delta_{\rm eff}^2\sigma_R^2S_4^{(2)}+12 \alpha
\mu_2^{(2)}+3 S_3^{(1)} \alpha \delta_{\rm eff}^2
\sigma^2_R-2\delta_{\rm eff}^2\sigma^2_R\alpha S_4^{(2)}\right) \right]
\ee 

To linear order in $\epsilon$ the effective action reads
\be
G(\delta_{\rm eff}) \approx {\delta_{\rm eff}^2 \over 2 
\sigma_R^2}\left\{1 - \frac{S_{3,R}}{3}\delta_{eff}\right\} \;.
\ee 
Note that this function is only convex for $\delta_{\rm
eff}<1/S_{3,R}^{(1)}$, so that our result can be consistently applied
only as long as $\delta_c(z_c)<1/S_{3,R}^{(1)}$.

We finally obtain
\begin{equation}
P(>\delta_c|z_c,R) \approx {1 \over \sqrt{2 \pi}} 
{\sigma_R \over \delta_c(z_c)} \exp\left[-\frac{1}{2}
{\delta_c^2(z_c) \over \sigma_R^2}
\left(1-\frac{S_{3,R}}{3} \delta_c(z_c) \right) \right]\;. 
\label{eq.pdeltac1O}
\end{equation}
Note that, for the Gaussian case, $S_{3,R}=0$, this formula corresponds
to the well-known asymptotic behaviour of the complementary error function
valid where $\delta_c \gg 2 \sigma$. In
particular, for $S_{3,R}=0$, and $z=0$ this approximation introduces
an error of $20\%$ at $R\sim10$ Mpc $h^{-1}$ and an error smaller than
$5\%$ for $R>20~h^{-1}$ Mpc.

For $\delta_c(z_c) \geq 1/S_{3,R}$, i.e. for $R \lap 10$ Mpc and/or
$\epsilon \gap 10^{-3}$ for model A, 
the integral  (\ref{eq.exactandgeneral}) has to be performed numerically.
To first order in $\epsilon$ eq. (\ref{eq.exactandgeneral}) becomes:
\begin{equation}
P(>\delta_c\mid z_c,R)=\frac{1}{2}-\frac{1}{\pi}\int_0^{\infty}
\frac{d\lambda}{\lambda}\exp(-\lambda^2\sigma_R^2/2)
\sin(\lambda\delta_c+\lambda^3\mu_{3,R}/6) \;. 
\label{eq:numeric}
\end{equation}
In Figure \ref{fig.Pdeltacnosp} we show the result of the numerical
integration for model A for masses in the range 
$10^8-10^{12} M_{\odot}$, redshift
$z_c=6,8,10$ and $\epsilon$ from top to bottom $\epsilon= 10^{-2}$,
$5 \times 10^{-3}$, $2 \times 10^{-3}$, $10^{-3}$, $5 \times 10^{-4}$,
and the Gaussian case ($\epsilon=0$, solid line). The main effect of
the presence of a small non-Gaussianity such as $\epsilon=5\times
10^{-4}$, is to amplify 
$P(>\delta_c\mid z_c,R)$ by a factor of order 10. 

For model B instead, as shown in Figure \ref{fig:sigmamuqpot},
the skewness parameter for
the density field is small and has the opposite sign of $\epsilon$. 
For this reason the
saddle-point approximation (60) works remarkably well on galaxy scales for
$|\epsilon|\lap 500$, if $z_c\lap 10$. 
Figure \ref{fig.Pdeltacnospmodelb} shows the excursion set
probability for model B for masses in the range 
$10^8-10^{12} M_{\odot}$, redshift
$z_c=6,8,10$ and $\epsilon$ from top to bottom $\epsilon=
-300,-200,-100,0,100,200$. The thick solid line is relative to the Gaussian
case ($\epsilon=0$); curves above that have $\epsilon<0$, below have
$\epsilon>0$. The fact that $|\epsilon|\sim 100$ is needed to
create any sizeable departure from the Gaussian $P(>\delta_c|M)$, has an
important consequence: all conventional inflationary models induce
deviations from Gaussianity such that the skewness parameter for the 
peculiar gravitational potential $S_{3,\Phi}$ is bound to be 
$|S_{3,\Phi}| \lap 10$ 
\citep{GLMM94}; since $S_{3,\Phi} \sim 6\epsilon$, we conclude that,
in the context of single-field inflation models the level-crossing 
probability $P(>\delta_c|M)$ 
is indistinguishable from the Gaussian one at least
on clusters scale and below. 
It is important to stress that for this model, using values of
$\epsilon$ of order unity or even larger does not imply the
breakdown of the perturbation expansion upon which our results rely,
as the coefficients of that expansion, such as the skewness parameter,
keep small even for $|\epsilon| \lap 100$. 



The most complex issue is that of finding a suitable expression for 
the function $P(M|M^\prime)$ in equation (51) in the general
non-Gaussian case and for generic
filters. In this context, it is useful to define a `fudge factor'
$f$ through the equation
\be
P(M|M^\prime) \equiv \frac{1}{f}~\Theta(M^\prime - M) \;,
\ee
with $\Theta$ the Heaviside step function. As shown by Nagashima \&
Gouda (1997),
in the Gaussian case, with sharp-k-space filtering, $P(M|M^\prime)$ 
reduces to the conditional probability 
$P(\delta_M\geq \delta_c|\delta_{M^\prime}=\delta_c)$, which
is easily obtained from a bivariate Gaussian, using
Bayes theorem; from this one immediately gets $f=2$, as expected (e.g. Peacock
\& Heavens 1990). In the
non-Gaussian case (and/or for different types of filter) 
the value of $f$ should be obtained from its very definition.
Recently, Koyama, Soda \& Taruya (1999) showed that in generic 
non-Gaussian models one can still write 
\be
f(M,M^\prime) \approx \left[\int_{\delta_c} d \omega 
~p(\delta_M=\omega~|\delta_{M^\prime}=\delta_c) \right]^{-1} \;,
\ee
with $p(\delta_M=\omega|\delta_{M^\prime} = \delta_c) \approx 
p(\delta_M=\omega-\delta_c)$, provided $M^\prime \gg M$.  
Obtaining a similar relation for {\it all} scales $M^\prime >M$ 
is a more complex issue, which would require a separate analysis 
\footnote{In the Gaussian case, and for sharp-k-space filter, one finds 
  $p(\delta_M=\omega|\delta_{M^\prime} = \delta_c) = 
  p(\delta_{\widetilde M}=\omega-\delta_c)$, for all $M^\prime >M$, where 
  ${\widetilde M}$ is defined by $\sigma^2({\widetilde M}) = \sigma^2(M) - 
  \sigma^2(M^\prime)$. Inserting this result in eq. (64) yields
  $f=2$, independently of the mass (Nagashima \& Gouda 1997).}.  
Nonetheless, it is extremely reasonable to expect that this  
approximation works well at least in so far as the deviation from
Gaussianity is weak. 
One then immediately gets the simple result \footnote{Note that,
if the non-Gaussian PDF depends on the filtering mass only through the
variance, i.e. $P(\delta_M)=\sigma_M^{-1} \Psi(\delta_M/\sigma_M)$, as assumed
by \citet{Willick99}, then $P(>0|M)=P(>\delta_c|M=0)$,
which yields an alternative expression for $f$, used by some authors 
(e.g. Robinson \& Baker 1999). In the most general case, however, the
latter result is not valid.} $f \approx 1/P(>0|M)$. 
In our case, one can compute the cumulative zero-crossing probability
$P(>0,M)$ numerically, starting from eq. (\ref{eq:numeric}).

The comoving mass-function of halos formed at
redshift $z_c$ can thus be immediately obtained by differentiating the
integral equation (51). This allows the standard PS formula to be extended to 
the non-Gaussian case in the simple form 
\citep{LM88,Colafrancescoetal89,COS97,RGS99a,RGS99b,KST99}: 
\begin{equation}
n(M,z_c) = f {3 H_0^2 \Omega_{0m} \over 8 \pi G M} \bigg| 
{dP(>\delta_c|z_c,R) 
\over d M} \bigg|  \;,
\end{equation}
where we implicitly assumed that the value of $f$ has negligible 
dependence on $M$,
The factor $f$ should account for the physical constraint that
all of the clustered matter in the Universe must be included in
bound objects of some mass [i.e. $\int_0^\infty dM M n(M,z_c) = 3H_0^2
\Omega_{0m}/8\pi G$] and simultaneously solve the cloud-in-cloud problem.

For the non-Gaussian models considered here, the
factor $f$ is not much different from the Gaussian value, provided 
$\epsilon$ is small, i.e. for small departures from the
Gaussian behaviour. In Figure \ref{fig.cic} the fudge factor $f\simeq
1/P(>0|M)$ is plotted as a function of the skewness parameter,
using a top-hat filter in real space.  
The useful range is on the left of the vertical dotted line, that is 
$M\gap 2\times 10^{10} M_{\odot}$ for model A with $\epsilon= 10^{-3}$ 
and $M\lap 4\times 10^{15}M_{\odot}$ for model B with $\epsilon =-100$.
We can  therefore conclude that, for mild non-Gaussianity, the  correction to 
the  usual fudge factor $f=2$ is negligible when compared to the effect of 
the skewness term in the exponential, justifying therefore the use of $f=2$ 
throughout. 

\subsection{Model A}

In model A the skewness parameter $S_{3,R}$ is scale-independent on
large scales,
therefore, as a first step, we can neglect the dependence of $S_{3,R}$ on mass.
In this case, in the above
differentiation, we arrive to a very simple generalization of the PS
formula, which has the same functional form, provided one makes 
the replacement 
\begin{equation}
\delta_c(z_c)  \to \delta_c(z_c) \left[1 - \frac{S_{3,R}}{3}
\delta_c(z_c) \right]^{1/2} \equiv \delta_{*}(z_c)  
\label{eq:largescalesubs}
\end{equation}
and therefore:
\be
n(M,z_c)\simeq f\frac{3 H_0^2\Omega_{0m}}{8 \pi G
M^2}\frac{\delta_{*}(z_c)} {\sqrt{2\pi}~\sigma_M}
\exp\left[ - \frac{\delta_{*}^2(z_c)}{2 \sigma^2_M}\right]
\bigg|\frac{d\ln\sigma_M}{d\ln M}
\bigg|
\label{eq:subst}
\ee
This  simple result is completely analogous to that found by
\cite{LM88}, 
for hierarchical-type statistics: the effect of the
deviation from primordial Gaussianity is that of shifting the 
PS characteristic mass to higher or lower masses,
depending on whether the skewness is positive or negative. Note that the 
effect of the non-Gaussian tail becomes more and more important at
higher redshifts. As noticed by \cite{Fry86}, `the departures from 
Gaussian can be appreciable while fluctuations are still small'. 
The  assumption that $S_{3,R}$ is scale-independent is correct 
on relatively large scales, where $R>20~h^{-1}$ Mpc,
corresponding to masses $M> 10^{14} M_{\odot}$.  Substitution
(\ref{eq:largescalesubs}) applied to the standard PS formalism agrees with the
numerical results on scales $R>20~h^{-1}$ Mpc, for $\epsilon \lap 10^{-2}$.
The dependence on the cosmological model is confined to $\delta_c(z_c)$, as
discussed before, once the present-day power-spectrum shape and
normalization have been chosen, and the normalization $H_0^2 \Omega_{0m}$.

However from Figure \ref{fig:sigmamuq} it is clear that the scale dependence
of $S_{3,R}^{(1)}$ is not negligible for galaxy masses, moreover 
for $R \lap 10$ i.e. $M \lap 10^{13}$ the saddle-point
approximation does not hold.
Numerical evaluation of (\ref{eq.exactandgeneral}) yields $dP(>\delta_c)/dM$;
the corresponding  $n(M,z_c)$ is shown in  Figure \ref{fig.nofm} for the same
range of masses, redshift of collapse $z_c$ and non-Gaussianity parameter
$\epsilon$ as in Figure \ref{fig.Pdeltacnosp}, assuming $f=2$, according 
to the discussion above. 


\subsection{Model B}
It is clear from the scale dependence of the skewness parameter that, 
for departures from Gaussianity of the kind of model B, 
clusters scales are better
than galaxy scales to probe primordial non-Gaussianity.
It is worth to notice here that also for model B in the absence of the
transfer function, $S_{3,R}$ is scale independent and indeed the skewness
parameter becomes scale-independent on very large scales ($R \gap 100
~h^{-1}$ Mpc), 
where the transfer function is unity. On clusters scales $S_{3,R}$
is not scale-independent, but for $\epsilon \lap 200$ and redshift $z_c\lap 2$
the saddle-point approximation is still valid.  For the comoving mass-function
of halos formed at redshift $z_c$ we obtain: 
\be 
n(M,z_c)\simeq f\frac{3H_0^2\Omega_{0m}}{8 \pi G M^2}\frac{1}
{\sqrt{2\pi}\sigma_M}\left|\frac{1}{2}\frac{\delta_c^2(z_c)}
{3\sqrt{1-S_{3,M}\delta_c(z_c)/3}} \frac{d S_{3,M}}{d \ln M}
+\frac{\delta_{*}(z_c)}{\sigma_M}\frac{d\ln \sigma_M}{d\ln M} \right|
\exp\left[ - \frac{\delta_{*}^2(z_c)}{2 \sigma^2_M}\right] 
\ee 
Also in this
case the dependence on the cosmological model is confined to $\delta_c(z_c)$
and in the normalization $H_0^2\Omega_{0m}$.  Figure \ref{fig.nofm-modelb}
illustrates the effect on the mass-function on clusters scales for the
non-Gaussianity of model B. For two different redshifts of collapse
($z_c=1$ and $2$) the ratio of the mass-function for model B to the 
mass-function for a Gaussian field [$n(M,z_c)/n_{\rm Gau}(M,z_c)$] 
is plotted as a
function of the mass in solar masses. Also in this case we can assume
$f=2$. The choice for the $\epsilon$ parameter
is, from top to bottom, $\epsilon=-100,-50,-10$.  It is clear that for high
masses one is probing the tail of the distribution, that is most sensitive to
departures from Gaussianity.  The forthcoming X-ray Multi-Mirror (XMM) galaxy
cluster survey \citep{RVLM99}, will allow the number density of clusters to be
 accurately measured even at $z\gap 1$. In particular the XMM cluster survey
will provide a complete survey of clusters  with masses $M \gap 10^{14}
M_{\odot}$ at $z<1.4$ over 800 square degrees. Preliminary calculations 
suggest that it will be therefore possible to detect $|\epsilon| \gap 50$.

\section{A worked example: galaxies at $z>5$}

The technique presented in the previous section allows one to measure
{\it accurately} the amount of non-Gaussianity on a given scale.
Although the traditional route has been to use the abundance of
clusters, we will illustrate our technique by using the observed
abundance of high-$z$ galaxies 
(e.g. Cavaliere \& Szalay 1986; Kashlinsky \& Jimenez 1997; Peacock
et~al. 1998). There are some advantages in
using high-$z$ galaxies: they sample directly the galaxy scale and the
objects are always inside virialised halos. One major disadvantage is
that the mass is not accurately determined. On the other hand, as
larger telescopes, such as NGST, get on line, it will be possible to
obtain volume limited samples of of high-$z$ galaxies and from
high-resolution spectra, dynamical masses will be determined more
accurately. As seen from eq. (\ref{eq:largescalesubs}), it is clear
that the higher the redshift the more sensitive the PDF is to
non-Gaussianity. Up to date there are 6 galaxies with spectroscopic
redshifts observed between $z>5$ to $z \approx 7$
\citep{DSSGC98,SSBDLYPF98,WSBSCTS98,HMC99,CLP99} and in Table~1 we
have compiled their most relevant characteristics. These high-$z$
galaxies have been found in relatively small areas of the sky: for the
purposes of this worked example we will use a scanned area of $3\times
10^{-4}$ square degrees for each galaxy \footnote{Four of the six
  galaxies considered here have been found in the Hubble Deep field,
  where the total area scanned is about $\sim$ 1/800 of a square
  degree. The other two galaxies have been found in Keck observations,
  where the area scanned for each of them is about $7.4 \times
  10^{-5}$ square degrees. For this worked example we will therefore
  make the conservative choice that there is on average one galaxy per
  $3\times 10^{-4}$ square degrees.}.

As pointed out before, the masses of this high-$z$ sample are unknown, but
their {\it current obscured} star formation rate is known. The first step is
to correct the observed star formation rate for the presence of dust. In
\citep{JPDBJM99} a case has been made for the correction factor for the star
formation due to the presence of dust in a starburst being $\sim 5$
(independently of the magnitude of the host galaxy). This is confirmed by the
independent arguments presented in \cite{Peacock+99}. The corrected star
formation rate is presented in Table~1. Since the age of the Universe at
$z\approx 6$ is already about 0.7 Gyr even for EdS, assuming the present age
to be 13 Gyr, it is very unlikely that galaxies have been observed just when
their stellar populations are born. Note that galaxies are likely to be
dust--enshrouded for about 0.02 Gyr \citep{JPDBJM99} and thus be invisible for
about 3\% of the Hubble time at $z\approx 6$. The star formation rate of the
galaxy in the past is more likely to be higher \citep{HJ99} than the present
observed one. To account for this we will then assume that the galaxy has been
observed when it is 0.1 Gyr old (i.e. about 10\% of the Hubble time at $z
\approx 6$) and has been forming stars at the current
(constant) observed rate. Using this, we can estimate what the mass in stars
in the galaxies (see Table~1) and in dark matter is, using a simple isothermal
profile for the halo, i.e. $M=V_c^3/(10GH(z_c))$, where $z_c$ is the redshift
of collapse \footnote{Although we have assumed that the stellar population of
the galaxy has to be about 10\% of the Hubble time at the observed redshift,
we will conservatively consider that the dark halo has just collapsed at the
observed redshift.  This is what is expected in hierarchical models, where the
stellar population is generally made in previous generations inside smaller
haloes.} and assuming that $\Omega_{0b}=0.015h^{-2}$. Following
\cite{PJDWSSDW98} we will carry out our analysis in terms of $\sigma_v$. For
an isothermal halo $\sigma_v=V_c/\sqrt{2}$. Table~1 shows that
$\sigma_v=70-200$ km s$^{-1}$. The number density of high-$z$ galaxies is
$N(\Omega=1) \gap 3.6 \times 10^{-3} (h^{-1} {\rm Mpc})^{-3}$
[N($\Omega_0=0.3,\Lambda_0=0.7$)$\gap 8.3 \times 10^{-4} (h^{-1} {\rm
Mpc})^{-3}$].

To compare these observations with our theoretical results we integrate the
mass-function to obtain the number of galaxies observed per unit volume with
redshift of formation $6 < z_c < 8$ and masses $2\times 10^{10}M_{\odot} < M <
4\times 10^{11} M_{\odot}$. It is worth noting that our mass estimates
slightly underpredict the masses that are derived using the width of the
$Ly\alpha$ line \citep{DSSGC98}. Gaussian initial conditions give
N($\Omega_0=0.3,\Lambda_0=0.7$)= $5.2\times 10^{-5}(h^{-1} {\rm Mpc})^{-3}$,
which is about a factor sixteen lower than the observed value. For 
Model A with $\epsilon = 10^{-3}$, we obtain N($\Omega_0=0.3,\Lambda_0=0.7$)=
$1.3\times 10^{-3}(h^{-1} {\rm Mpc})^{-3}$, which is much closer to the
observed value. As we already noted before, a very small non-Gaussianity
parameter (i.e. a very small departure from the Gaussian behaviour) 
has a dramatic effect on the statistics of high redshift objects. This
is, of course, very preliminary. 
For the tail of the PDF, the mass function is very sensitive to the
overdensity threshold, the mass determination and the redshift of
formation. 
In this example we have assumed that the redshift of formation is the
redshift at which the object has been observed. This is of course quite a
conservative assumption, yielding a lower limit on the non-Gaussianity
parameter.  It has been shown that the PS algorithm with a fixed
threshold $\delta_c$ does not reproduce very accurately N-body simulation
results (e.g. Sheth \& Tormen 1999). However, the
deviations on the mass scales considered here are never larger than a
factor of a few, well below the uncertainty due to the mass determination
(see below). PS predictions and N-body results can be reconciled by making
the threshold for collapse scale-dependent (e.g. Sheth \& Tormen 1999;
Sheth, Mo \& Tormen 1999). This will
change the absolute prediction on the number density of objects but will
not strongly affect the ratio between the Gaussian and non-Gaussian
number density predictions. 
The spherical collapse assumption underlying the PS algorithm is the main reason why theory does not reproduce accurately N-body results: the
introduction of the ellipsoidal collapse improve sensibly the agreement (Lee
\& Shandarin 1998).
In a subsequent paper we will address the issue of triaxial collapse for
non-Gaussian initial conditions.

Maybe the most pressing issue is that of the mass determination of
high-redshift objects. For example, if the mass estimate for the galaxies
in the worked example is wrong by a factor of 4, the number density would
change by a factor of 20 (at $z_c=6$), and agreement between predictions
and observations could be obtained without resorting to $\epsilon \ne 0$.

Indeed, better observations will improve the determination of the
mass of the dark halos and of the number density of high redshift galaxies.  A
measurement of $\epsilon$ and thus a determination of the amount of
primordial non-Gaussianity (if any!) present on galaxy scales will then
become possible.

\section{Conclusions}

We have shown that it is possible to test whether initial conditions were
Gaussian, by looking at the abundance of galaxies and clusters at high
redshift.  Small deviation from Gaussianity have a deep impact on those
statistics which probe the tail of the distribution: for this reason the
abundance of high redshift objects such as galaxies at $z\gap 5$ and clusters
at $z\gap 1$, is very sensitive to small departures from the Gaussian 
behaviour. 

We used a parameterization of primordial non-Gaussianity that covers a
wide range of physically motivated models: the non-Gaussian field is
given by a Gaussian field plus a term proportional to the square of a
Gaussian field, the proportionality constant being the non-Gaussianity
parameter. We applied this model to the primordial density fluctuation
field and to the gravitational potential fluctuation field, and we
assumed that the departures from the Gaussian behaviour are small.
Given this parameterization of non-Gaussianity, we were able to obtain
an analytic expression for the probability density function (PDF) of
the smoothed density field. This approach is somewhat different from
what is found in the literature (e.g. Robinson et~al. 1999a,b), since we
do not assume a particular form for the non-Gaussian PDF, but we {\it
derive} it, given a parameterization of the primordial
non-Gaussianity.

We then introduced a generalized version of the PS approach valid in
the context of non-Gaussian initial conditions, also tackling the
cloud-in-cloud problem. This enabled us to relate, analytically, the
non-Gaussianity parameter to the number of high-redshift objects. The
strength of our method is that we are able to properly take into
account the smoothing scale, or mass, dependence of the probability
distribution function, and that our results are analytic.  In looking
at the likelihood of rare events for a non-Gaussian density field, one
is probing the tail of the distribution, and analytical results are
not only elegant but also extremely important.

The main technical results of the present paper can be summarized as
follows: a) Eq. (27) is the exact analytic form of the
cumulant generator for our non-Gaussian models: it 
can be used to generate cumulants of any order by simple
differentiation.  b) Eq. (\ref{eq.exactandgeneral}): this equation gives an
exact and general analytic expression for the cumulative probability
that, at a given redshift $z_c$, the fluctuations on scale $R$ exceed some
critical value $\delta_c$. From this expression it is
possible to obtain the cumulative probability to any order by expanding the
cumulant generator (36) with (40) and (41) as we did in (50). c)
Eq. (61): it is an analytic expression for the cumulative 
probability  valid for
small deviations from the Gaussian behaviour.  d) Eq. (68), with the
definition (66): it is an analytic expression for the comoving mass-function
of halos obtained within the PS approach. For a given power spectrum
the dependence on the cosmological model is straightforward: it is
confined to the normalization factor $H_0^2\Omega_{0m}$ and the threshold
$\delta_c$.  

Still an important issue remains unsolved: how to
accurately determine the mass and the formation redshift of observed
high-redshift objects in order to put tight constraints on primordial
non-Gaussianity.  This issue will be addressed in a forthcoming paper.

\acknowledgments 
LV acknowledges the support of a TMR grant. LV and RJ thank
A. Berera, A. Heavens and J. Peacock for stimulating discussions and the
anonymous referee for useful comments. SM thanks
K. Jedamzik, H. J. Mo and R. Sheth for useful discussions.


\clearpage

\noindent
{\bf Figure 1:} {\bf Model A} variance, skewness (\mbox{$\mu_{3,R}^{(1)}$})
and skewness parameter ($S_{3,R}^{(1)}$) as a function of the radius of the
top-hat window in real space and of the mass. To produce this plot we assumed:
$\alpha=1$, a scale-invariant initial power-spectrum $P_{\phi} \propto k$, CDM
transfer function, $\Omega_{0m}=0.3$, normalized to produce $\sigma_8(z=0)=1$
(more details in the text).

\noindent
{\bf Figure 2:} {\bf Model B} variance, skewness (\mbox{$\mu_{3,R}^{(1)}$})
and skewness parameter ($S_{3,R}^{(1)}$) as a function of the radius of the
top-hat window in real space and of the mass. To produce this plot we made the
same assumptions as for Figure 1 except that now $P_{\phi}\propto k^{-3}$.

\noindent
{\bf Figure 3:} Left panel: The redshift dependence of the threshold for
collapse $\Delta_c$ in different cosmologies. It is clear that the dependence
on redshift and on cosmology is weak. This justifies to assume
$\Delta_c$=1.687 throughout. Middle panel: Dependence of the relation between
dark mass and radius on the cosmology and the redshift of collapse $z_c$. The
solid lines denote an Einstein de Sitter Universe, the dashed lines a flat
model with $\Omega_{0m}=0.3$ and $\Lambda_0=0.7$ and the dotted lines an open
model with $\Omega_{0m}=0.3$. The dependence is very weak and can be safely
ignored. Right panel: The redshift dependence of the inverse of the linear
fluctuation growth for different cosmologies.

\noindent
{\bf Figure 4:} {\bf Model A} $P(>\delta_c\mid z_c,R)$ for $z_c=6$ (left
panel), $z_c=8$ (center), $z_c=10$ (right panel) and $\epsilon=10^{-2}$, $5
\times 10^{-3}$, $2\times 10^{-3}$, $10^{-3}$, $5 \times 10^{-4}$.  The thick
solid line is relative to the Gaussian case. It is clear that the main effect
of the presence of a small nonzero $\epsilon$ is to boost $P(>\delta_c\mid
z_c,R)$ by at least a factor $\sim 10$.

\noindent
{\bf Figure 5:} {\bf Model B} $P(>\delta_c\mid z_c,R)$ for
$z_c=6$ (left panel), $z_c=8$ (center), $z_c=10$ (right panel) and
$\epsilon=-300$, $-200$, $-100$, $0$, $100$, $300$ (from top to bottom). The
thick solid line is relative to the Gaussian case. It is clear that
$\mid\epsilon_B\mid$ of model B need to be $\gg \epsilon_A$ of model A
to show a noticeable
difference from the Gaussian case. Moreover $\epsilon_B<0$ is needed in order
to enhance the non-Gaussian tail of $P(>\delta_c\mid z_c,R)$.

\begin{figure}
\begin{center}
\setlength{\unitlength}{1mm}
\begin{picture}(140,55)
\includegraphics{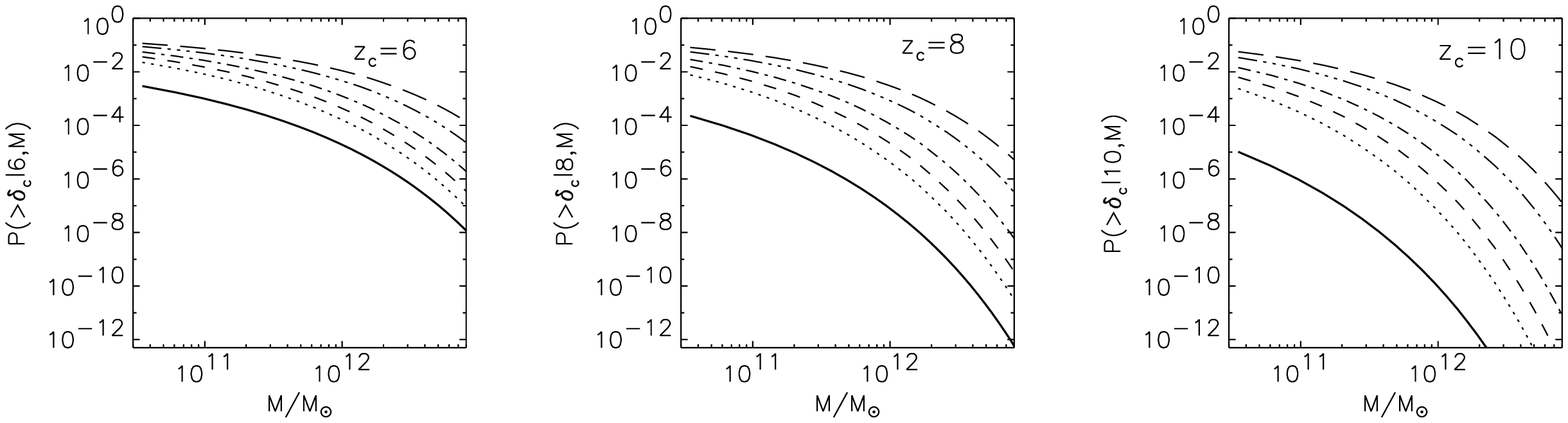}
\end{picture}
\end{center}
\caption{}
\label{fig.Pdeltacnosp}
\end{figure}

\begin{figure}
\begin{center}
\setlength{\unitlength}{1mm}
\begin{picture}(140,55)
\includegraphics{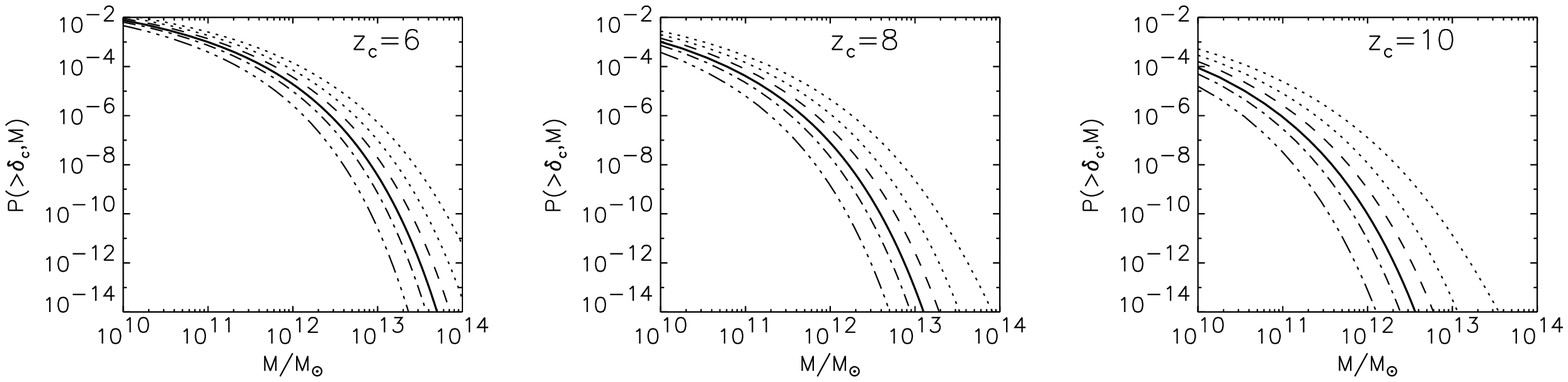}
\end{picture}
\end{center}
\caption{}
\label{fig.Pdeltacnospmodelb}
\end{figure}

\begin{figure}
\begin{center}
\setlength{\unitlength}{1mm}
\begin{picture}(140,55)
\includegraphics{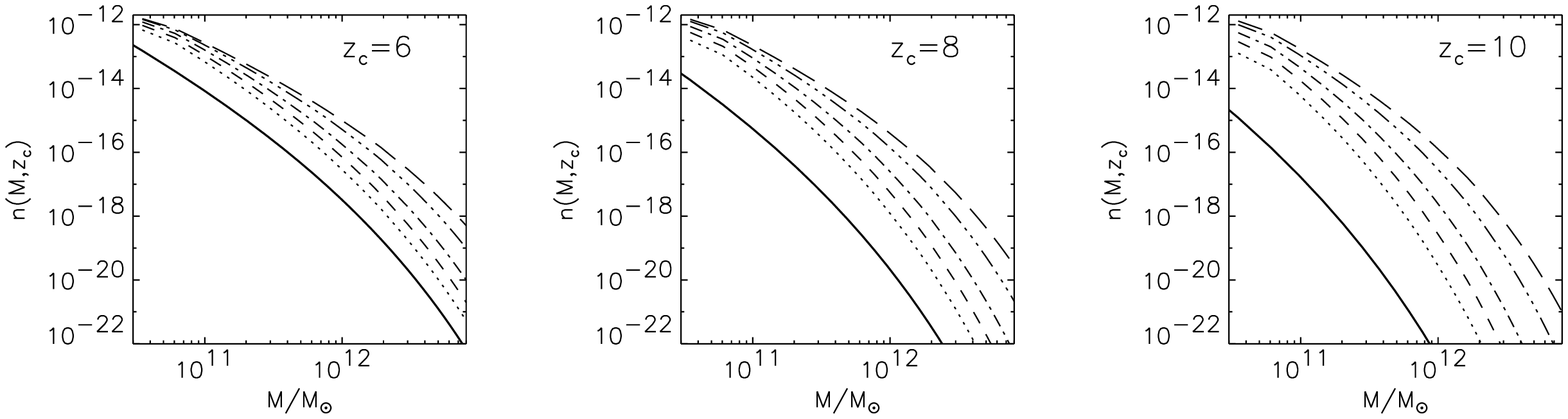}
\end{picture}
\end{center}
\caption{}
\label{fig.nofm}
\end{figure}

\begin{figure}
\begin{center}
\setlength{\unitlength}{1mm}
\begin{picture}(140,55)
\includegraphics{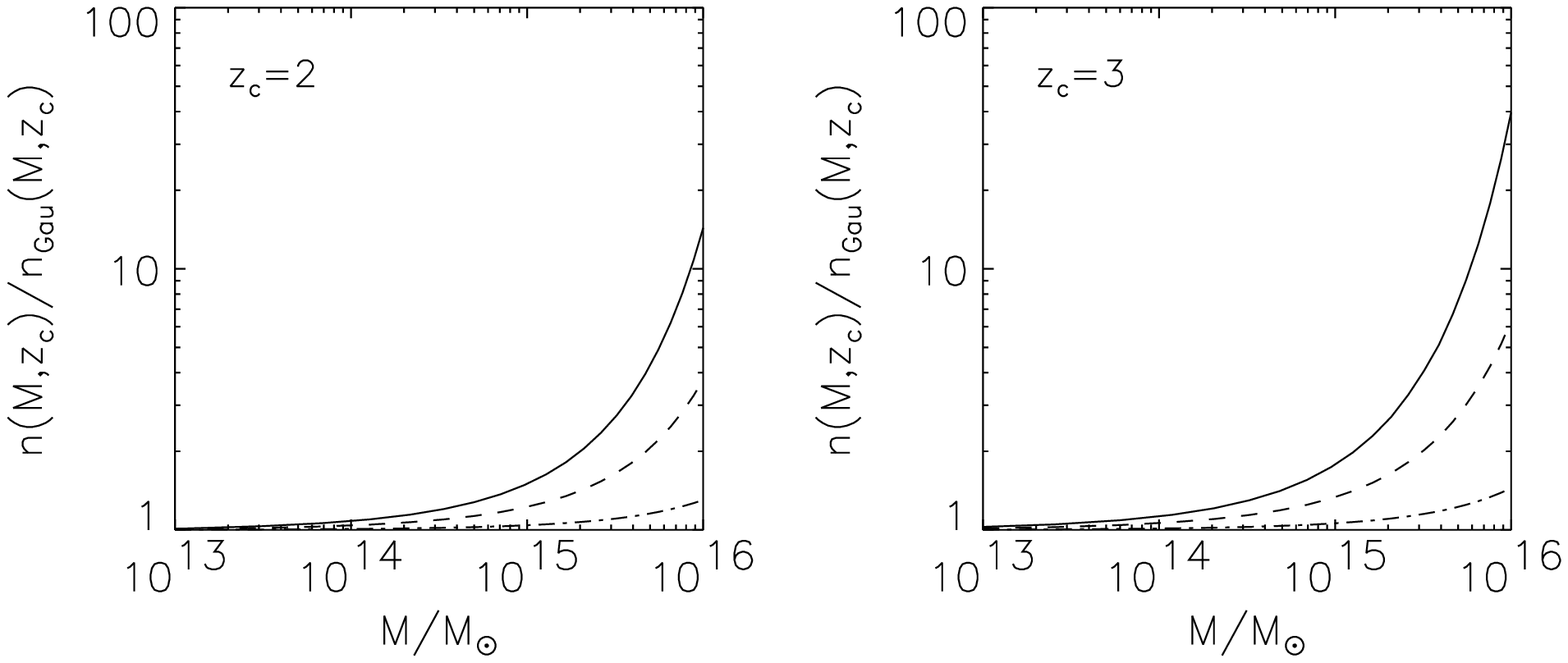}
\end{picture}
\end{center}
\caption{}
\label{fig.nofm-modelb}
\end{figure}

\begin{figure}
\begin{center}
\setlength{\unitlength}{1mm}
\begin{picture}(140,55)
\includegraphics{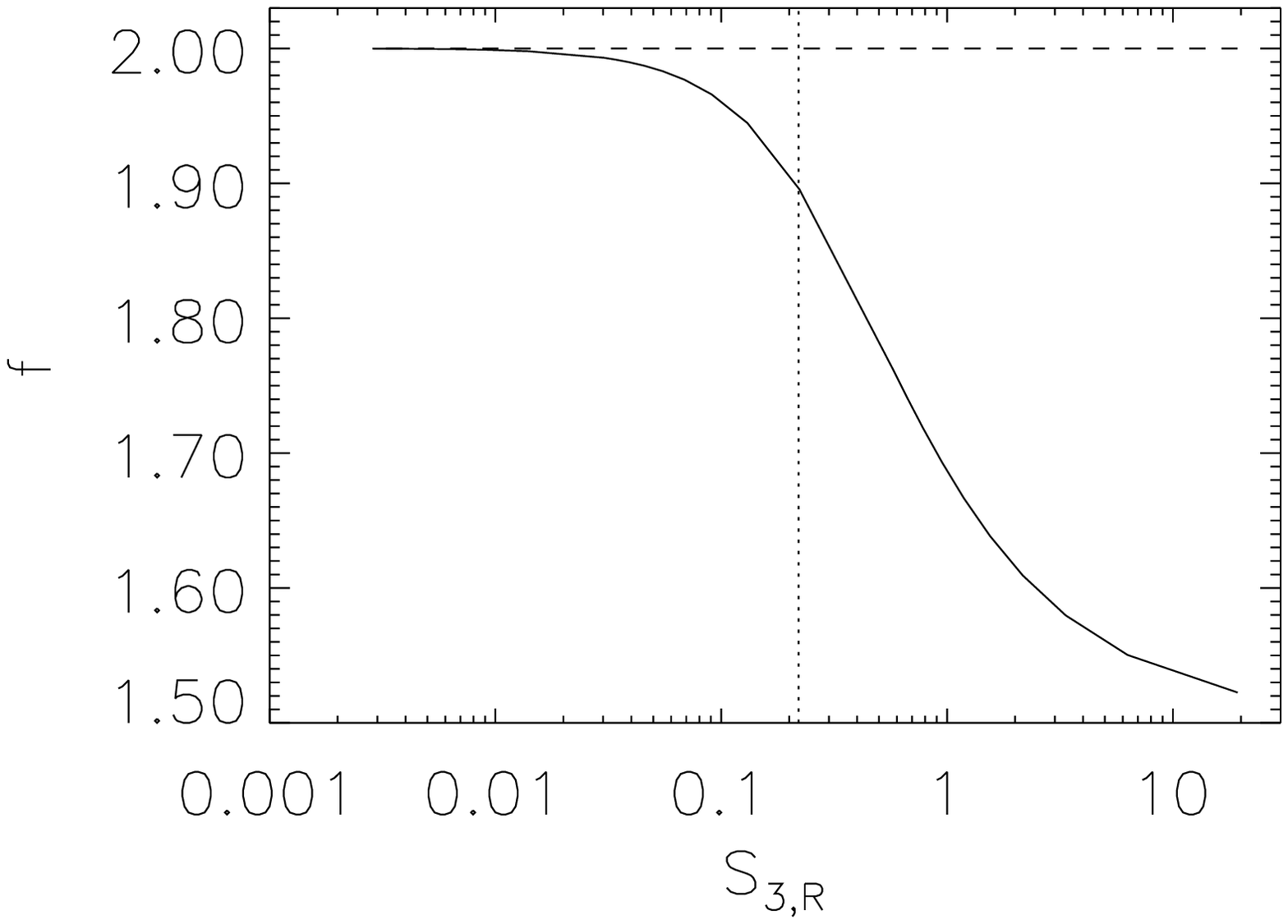}
\end{picture}
\end{center}
\caption{}
\label{fig.cic}
\end{figure}

\noindent
{\bf Figure 6:} {\bf Model A}: The comoving mass-function of halos formed at
redshift 6 (left panel) 8 (center), 10 (right panel), for different values of
the non-Gaussian parameter $\epsilon$, as in Figure \ref{fig.Pdeltacnosp}; from
top to bottom: $\epsilon=10^{-2}$, $5\times 10^{-3}$, $2 \times 10^{-3}$,
$10^{-3}$, $5 \times 10^{-4}$. The thick solid line is relative to the
Gaussian case.

\noindent
{\bf Figure 7:} {\bf Model B}: The non-Gaussian effect on the
mass-function on clusters
scales. For two different redshift of collapse ($z_c=1$ left and $z_c=2$
right) the ratio of the mass-function for {\bf model B} to the mass-function
for a Gaussian field [$n(M,z_c)/n_{\rm Gau}(M,z_c)$] is plotted as a function
of mass. The choice for the $\epsilon$ parameter is, from top to bottom,
$\epsilon=-100,-50,-10$.  It is clear that for high masses one is probing the
tail of the distribution, that is most sensitive to departures from
Gaussianity.

\noindent
{\bf Figure 8:} The fudge factor $f\simeq 1/P(>0|M)$ as a function of the 
skewness parameter. The value $f=2 $ is exact for the Gaussian case
with sharp-k-space filter; the solid line shows the value of $f$
as a function of the skewness parameter $S_{3,R}$. The useful range is
on the left of the vertical dotted line, i.e. $M\gap 2\times 10^{10}
M_{\odot}$ for model A with $\epsilon= 10^{-3}$ and $M\lap 4\times
10^{15}M_{\odot}$ for model B with $\epsilon =-100$.
We can conclude that, for mild non-Gaussianity, the  correction to the usual
fudge factor is negligible, justifying therefore the use of $f=2$ throughout.  

\clearpage

\begin{table}
\begin{center}
\caption{Main properties of observed galaxies at $z>5$.  Column 1: [1]
  Dey et al (1998);
  [2] Spinrad et al. (1998); [3] Weymann et al. (1998); [4] Hu, McMahon
  \& Cowie (1999); [5] Chen, Lanzetta \& Pascarelle (1999). Column 2:
  spectroscopic redshift of the galaxy. Column 3: area of the sky (in
  square degrees) surveyed per galaxy. Comumn 4: observed star
  formation rate in solar masses per year (assuming $h=0.65$). Column 5: dust corrected
  star formation rate (see text). Columns 6, 7: estimated mass in stars
  and dark matter (see text).  Column 8: velocity dispersion obtained
  assuming an isothermal halo (see text).}
\begin{tabular}{llllllll}
\hline
\hline
Ref. & 
$z$ & 
deg$^2$ &
 $\frac{\rm obs. SFR}{(M_{\odot} {\rm yr}^{-1})}$ & 
 $\frac{\rm corr. SFR}{({\rm M}_{\odot} {\rm yr}^{-1})}$ & 
$\frac{{\rm est.} M_*}{M_{\odot}}$ & 
$\frac{ {\rm est.} M_{\rm dark}}{M_{\odot}}$ & 
$\frac{\sigma_v}{({\rm km} {\rm s}^{-1})}$ \\
\tableline
$[1]$ & 5.34 & $7.4\times 10^{-5}$ & 4 & 20 &   2 $\times 10^9$ &   2 $\times 10^{10}$ & 70 \\
$[2]$ & 5.34 & $3\times 10^{-4}$ & 13 & 65 & 6.5 $\times 10^9$ & 6.5 $\times 10^{10}$ & 100 \\
$[2]$ & 5.34 & $3\times 10^{-4}$ & 13 & 65 & 6.5 $\times 10^9$ & 6.5 $\times 10^{10}$ & 100 \\
$[3]$ & 5.60 & $3\times 10^{-4}$ & 8 &  40 & 4$\times 10^{9}$ & 4 $\times 10^{10}$ & 86 \\ 
$[4]$ & 5.74 & $7.4\times 10^{-5}$ & 40 & 200 & 4$\times 10^{10}$ & 4$ \times 10^{11}$ & 187 \\  
$[5]$ & 6.68 & $3\times 10^{-4}$ & 40 & 200 & 4$\times 10^{10}$ & 4 $\times 10^{11}$ & 200 \\  
\tableline
\end{tabular}
\end{center}
\end{table}

\clearpage

\clearpage

\clearpage

\clearpage

\clearpage

\clearpage

\clearpage

\clearpage
\end{document}